\documentclass[floatfix, preprintnumbers, amsmath, amssymb, runinaddress, tightenlines, nofootinbib, 11pt]{revtex4}
\pdfoutput=1
\usepackage{epsfig,bbm,bm,amssymb,graphicx}

\def\be{\begin{equation}} 
\def\ee{\end{equation}} 
\def\bea{\begin{eqnarray}} 
\def\eea{\end{eqnarray}} 

\begin{document}

\newcommand{\eq}[2]{\begin{equation}\label{#1}{#2}\end{equation}}
\newcommand{\p}{\varphi}
\newcommand{\h}{\mathcal H}
\newcommand{\N}{\mathcal N}
\newcommand{\tbeta}{\tilde{\beta}}
\newcommand{\calH}{{\cal H}}
\newcommand{\calR}{{\cal R}}
\newcommand{\abs}{|}
\def\nn{\nonumber}

\date{\today}

\title{\Large{Cosmological Perturbations Across an S-brane}}

\author{Robert H. Brandenberger$^1$, Costas Kounnas$^2$,\\ Herv\'e Partouche$^3$, Subodh P. Patil$^{4}$ and Nicolaos Toumbas$^5$} 
\affiliation{1) Department of Physics, McGill University, 
Montr\'eal, QC, H3A 2T8, Canada\\}
\affiliation{2) Laboratoire de Physique Th\'eorique, Ecole Normale Sup\'erieure, 24 rue Lhomond, F-75231 Paris cedex 05, France\\}
\affiliation{3) Centre de Physique Th\'eorique, Ecole Polytechnique, F-91128 Palaiseau cedex,  France\\}
\affiliation{4) Theory Division, PH-TH Case C01600, CERN, CH-1211 Geneva, Switzerland\\}
\affiliation{5) Department of Physics, University of Cyprus, Nicosia 1678, Cyprus\vspace{5pt}}


\rightline{\footnotesize{CERN-PH-TH/2013-289}}
\rightline{\footnotesize{CPHT-RR001.0113}}
\rightline{\footnotesize{LPTENS-13/18}}

\begin{abstract}

\vskip .5cm


Space-filling S-branes can mediate a transition between a contracting and an expanding universe in the Einstein frame. Following up on previous work that uncovered such bouncing solutions in the context of weakly coupled thermal configurations of a certain class of type II superstrings, we set up here the formalism in which we can study the evolution of metric fluctuations across such an S-brane. Our work shows that the specific nature of the S-brane, which is sourced by non-trivial massless thermal string states and appears when the universe reaches a maximal critical temperature, allows for a scale invariant spectrum of curvature fluctuations to manifest at late times via a stringy realization of the matter bounce scenario. The finite energy density at the transition from contraction to expansion provides calculational control over the propagation of the curvature perturbations through the bounce, furnishing a working proof of concept that such a stringy universe can result in viable late time cosmology.   
\end{abstract}
\maketitle

\section{Introduction}
Since its inception, the inflationary universe scenario \cite{Guth} has widely come to be accepted as one of the most promising phenomenological models of the early universe that successfully accounts for the initial conditions of the hot Big Bang\footnote{We remark that operationally speaking, this means that among the parameters of the six parameter $\Lambda-$CDM model which purports to account for most cosmological observations, the simplest models of inflation typically account for one (the spectral tilt) and post-fit another (the amplitude of the primordial power spectrum). In contrast, in certain examples of putative models where the universe bounces or is emergent, the amplitude is also fixed by the characteristic scales of the underlying dynamics \cite{NBV, JLPS}.}. However, the paradigm is not without its conceptual \cite{RHBrev0, Turokrev} as well as technical problems\footnote{A subset of which includes inflation's sensitivity to its UV completion (see \cite{BG} for a discussion of the pertinent issues). The so-called ``eta" problem is but one facet of this, and although one can claim to have addressed this up to some subset of corrections, the problem is expected to persist in all generality \cite{leiden}.}, and thus it is of more than just academic interest
to study the possibility of alternative cosmological scenarios. Notable attempts motivated by string or M-theory towards this end include \cite{BV,TV,Veneziano,NBV,JLPS,SGCrev,Ekp}. 

Recently, there have been some developments in a specific class of non-singular\footnote{By non-singular in the context of these works and what follows here, we mean that the energy density (i.e. $H^2$) remains finite in the Einstein frame, although the extrinsic curvature does jump, sourced by a distributional source.} bouncing string cosmologies \cite{KPT,KPT2,FKPT,CorfuLectures}, and in earlier emergent scenarios, see e.g. \cite{RHBrev1}.  With respect to the former, it has been realized \cite{Wands, Fabio} that initial quantum vacuum fluctuations that exit the Hubble radius during a matter-dominated contracting phase acquire a scale-invariant spectrum of curvature perturbations on super-Hubble scales, representing a \textit{duality} of backgrounds that generates the same two point correlation functions \cite{Wands}\footnote{Concerning emergent scenarios (by which we mean scenarios in which there is an initial quasi-static period as postulated in ``string gas cosmology" \cite{BV, SGCrev}, see also \cite{KPT2}), it has been realized \cite{NBV} that thermal string gas perturbations on a toroidal spatial background yield a scale-invariant spectrum of curvature fluctuations.}$^,$\footnote{These two scenarios hence become possible alternatives to cosmological inflation for producing the observed inhomogeneities in the distribution of galaxies and anisotropies in the cosmic microwave background temperature maps. The first, the ``matter bounce" scenario, has as a distinctive prediction a special shape of the bispectrum \cite{bispectrum}; the ``string gas cosmology" scenario predicts a slight blue tilt of the spectrum of primordial gravitational waves, whereas standard inflation predicts a red tilt \cite{BNPV}.}.

In order to propagate the initial spectrum of fluctuations in the contracting or emergent phase on to the final expanding phase, it is necessary to understand their evolution at the transition point. Note that this transition must involve new physics which violates the Null Energy Condition (NEC). In the context of bouncing universes, the matching of fluctuations between the phases was studied in a number of specific models in which the bounce was obtained by introducing ghost matter \cite{quintom} (where stability issues for the background are not considered) and ghost condensate \cite{condensate} or Galileon \cite{Galileon} constructions (where meta-stable violations of the NEC are possible). Additionally, other models consider modifying gravity at high curvature scales such as in the non-local model of \cite{Biswas}, in Ho\v{r}ava-Lifshitz gravity \cite{HLbounce} or in mirage cosmology \cite{Omid}.

In all of these examples it was found that on large scales (length scales
larger than the time-scale when the new physics is dominant)
the spectrum of the dominant mode of fluctuations is unchanged
across the bounce. On the other hand, it is known from studies
done in different contexts \cite{MP, DV} that the transition can
depend sensitively on details of the background bouncing solution.

In this paper, we study the transition of fluctuations between
the contracting and expanding phases in cosmological backgrounds arising from specific string theoretic constructions \cite{KPT,KPT2,FKPT,CorfuLectures} in which the bounce is mediated via a stringy S-brane. Our results show that indeed a scale invariant spectrum of curvature perturbations can manifest at late times, given vacuum initial conditions, in a stringy realization of the ``matter bounce" scenario \cite{Wands}.  The detailed manner however in which this happens crucially depends on the microphysical realization of the S-brane as a space-filling condensate of non-trivial, massless thermal string states, appearing at the instant the universe attains a critical maximal temperature, reinforcing the message of \cite{MP, DV}. Our work thus lays the basis for studying the evolution of cosmological fluctuations in general models in which the transition between the phases of contraction and expansion is described by stringy S-branes.

Specifically, we will study scalar metric fluctuations (see \cite{Mukhanovbook, MFB} for an in-depth survey of the theory of cosmological fluctuations and \cite{RHBrev2} for an introductory overview). In both the contracting and expanding phases of the induced cosmology, the gravitational potential $\Phi$ obeys an equation of motion which for each Fourier mode, labeled by co-moving wavenumber $k$, is a second order ordinary differential equation and hence has two linearly independent solutions. In the expanding phase, the dominant mode is constant on super-Hubble scales\footnote{This
is the case as long as there are no initial isocurvature perturbations.} and the second mode is decaying in time. In the contracting phase there is a growing mode in time and a second constant mode.

In the case of the toy bouncing models mentioned above, it
is found that on scales much larger than the duration of
the ``new physics phase", the coupling between the growing
mode in the contracting phase and the dominant mode in 
the expanding phase is highly suppressed, leading to the
conclusion that the spectrum of $\Phi$ at late times in
the expanding phase is determined by the spectrum of
the constant mode in the contracting phase. As shown by
Durrer and Vernizzi \cite{DV} (see also \cite{CW}), in the case of an instantaneous transition between contraction and expansion, the 
generalization of the Israel junction conditions \cite{Israel}
to spacelike hypersurfaces \cite{HV, DM} implies that the coupling of the 
growing mode in the contracting phase to the constant mode in the expanding 
period is {\it not} suppressed unless the matching surface is a constant
energy density surface. This result had important implications for
the transfer of fluctuations through examples such as an Ekpyrotic bounce. 
Simple four-dimensional single matter component effective field theory analyses assumed a
matching on constant energy density slices \cite{Lyth, Fabio2, Hwang}
and obtained a deeply blue spectrum of final curvature fluctuations starting 
from vacuum perturbations in the contracting Ekpyrotic phase\footnote{On the other hand, analyses taking the higher spacetime dimensional 
origin of the Ekpyrotic scenario into account \cite{Tolley, Thorsten} showed that the transition surface was not a constant energy density surface which led to the conclusion that the final spectrum of curvature fluctuations was
indeed scale-invariant, as argued initially in \cite{Ekp}. See also \cite{EkpNew} for recent discussions on the quantum nature of the fluctuations in the Ekpyrotic models.}.

By contrast, in this paper we study the cosmology induced in a concrete string theoretic construction \cite{KPT,KPT2,FKPT,CorfuLectures} in which an S-brane mediates the transition between the contracting and expanding phases. In these models, the S-brane is {\it not} located on a hypersurface of constant energy density. Instead, it sits on a  hypersurface of constant critical string-frame temperature $\tilde T\equiv \tilde T_c$ at which point a marginal operator on the worldsheet becomes relevant and induces direct transitions between thermal winding and thermal momentum states, thus interpolating between two different geometric phases of the underlying worldsheet CFT. In the Einstein frame the hypersurface is {\it not} isothermal since the temperature $T$ is dressed by the dilaton field: 
$T=\tilde T \,e^{\phi}=\tilde T\,g_{s}$, where $g_{s}$ is the string coupling. This yields to a new kind of hypersurface where the combination $Te^{-\phi}=\tilde T$ has a maximal and constant value $\tilde T_c$. As we will shall see further, the precise properties of this isothermal hypersurface yield an unsuppressed mixing between the spectrum of fluctuations of the growing mode of $\Phi$ in the contracting phase with that of the constant mode of $\Phi$ in the expanding phase, which translates at late times into a scale invariant spectrum for the curvature perturbation $\zeta$.  

The role of the S-brane is to glue the geometrical phases associated with the two dual asymptotically cold regimes in a consistent way in the low energy effective theory, and as such, provides the means to \textit{instantaneously} violate the NEC \cite{KPT,KPT2,FKPT,CorfuLectures}. The instantaneous nature of this violation simply does not permit the time for any gradient instabilities to set in -- the microphysical realization of the S-brane as a space-filling defect that mediates the phase transition between distinct geometric phases of the explicit underlying string construction guaranteeing its consistency \cite{KPT,KPT2,FKPT,CorfuLectures}.

In what follows, we begin by reviewing the background string construction and its low energy description at length, after which we consider perturbations of the system in longitudinal gauge (see Appendix \ref{adm4} for a discussion of other gauge choices). We then consider matching the perturbations across the S-brane, demonstrating that there is no thermal entropy production by the fluctuations of the S-brane.  
We find that due to the specific properties of the stringy S-brane, no isocurvature component is generated at the bounce and a scale invariant spectrum for the curvature perturbation results at late times. After contextualizing this calculation in the matter bounce scenario, we offer a summary and our concluding thoughts, taking care to highlight the open issues confronting our model. 

\section{Background}

\subsection{Thermal duality and the origin of the stringy S-brane}

We consider bouncing string cosmologies that are induced by specific weakly-coupled, thermal configurations of ${\cal{N}}=(4,0)$ superstrings \cite{KPT,KPT2,FKPT,CorfuLectures} compactified to four dimensions.  
Compactifications with less amount of (initial) supersymmetry 
could also be considered without changing the main conclusions concerning the evolution through the bounce. In order to study the thermodynamic properties of the system, we employ the Euclidean description where the time direction is compactified on a circle of radius $R$. In addition to temperature, non-trivial ``gravito-magnetic'' fluxes thread the Euclidean time circle, as well as internal spatial cycles responsible for the (spontaneous) breaking of the right-moving supersymmetries \cite{KPT, KPT2, FKPT,CorfuLectures,AKPT, K,DLS}. Effectively the fluxes suppress the exponentially growing density of thermally excited oscillator states, 
curing the Hagedorn instabilities of the canonical ensemble and restoring 
the T-duality symmetry along the Euclidean time direction. The one-loop string partition function is finite for all values of the thermal modulus $R$, and it is invariant under  thermal duality:
\be
Z(R) \, = \, Z\left(\frac{R_c^2}{R}\right) \, ,
\ee
where the self dual, critical radius $R_c$ is of order the string scale. 
In a large class of type II superstring models, the critical radius attains a universal value, $R_c^2=\alpha'/2$ (corresponding to the so-called fermionic point), irrespective of the dimensionality of spacetime.

A direct consequence of thermal duality is that the stringy system exhibits two asymptotically cold regimes, at $R\gg R_c$ and at $R\ll R_c$, dominated by the light thermal momentum and light thermal winding modes respectively. Each asymptotic regime gives rise to a local low energy effective field theory description, associated with a distinct $\alpha^\prime$-expansion, with the inverse temperature $\tilde\beta(R)$ being proportional to the period of the Euclidean time circle or its dual respectively:
\be
\label{beta(R)}
\tilde\beta = 2\pi R\;\;\mbox{  as  }\; \;R\gg R_c\, , \quad \tilde\beta =2\pi\,  
{R_c^2\over R}\;\;\mbox{  as  }\; \;R\ll R_c\, ,
\ee
where the tilde denotes quantities in the string frame. (Further on, untilded variables will denote Einstein frame quantities.) As a result the temperature in these models never exceeds a certain critical value, 
$\tilde T_c = 1/\tilde\beta(R_c)\sim 1/2\pi R_c$, which the system attains at the self-dual fermionic  point.

In the regime of light thermal momenta, $R \gg R_c$, the partition function is dominated by the thermally excited massless states and is given by \cite{KPT}
\be
\frac{Z}{V} \, = \, {n^{*} \sigma_r \over (2\pi R)^3} + {\cal{O}}(e^{- R/ R_c}) \, ,
\ee
where $V$ is the volume of our three-dimensional spatial torus, which we can
take to be arbitrarily large; $n^{*}$ is the number of the effectively massless degrees of freedom and $\sigma_r$ is Boltzmann's radiation constant. In the dual regime of light thermal windings, $R \ll R_c$, the expression for the partition function becomes
\be
\frac{Z}{V} \, = \, {n^{*} \sigma_r R^3\over (2\pi R_c^2)^3} + {\cal{O}}( e^{- R_c / R_0} ) \, .
\ee
In terms of the duality invariant inverse temperature $\tilde \beta$, Eq. (\ref{beta(R)}), the partition function in {\it both} asymptotic regimes is given by
\be
\frac{Z}{V} \, = \, {n^{*} \sigma_r \over {\tilde \beta}^3} + {\cal{O}}(e^{- \tilde \beta / \tilde \beta_c}) \, ,
\ee
leading to the characteristic equation of state of massless thermal radiation. The presence of 
asymptotic supersymmetry in the right-moving string
sector ensures that the partition function, and so the resulting equation of state,
can be well approximated with that of massless thermal radiation up to temperatures 
close to the critical one \cite{KPT}. In the neighborhood of the self-dual point, 
the one-loop thermodynamical quantities acquire stringy corrections\footnote{In the two dimensional hybrid models of \cite{FKPT} the equation of state is 
exactly that of massless thermal radiation up to the critical temperature thanks to 
exact right-moving massive spectrum boson/fermion degeneracy symmetry \cite{K}.}, 
including the precise relation between the inverse temperature and the radius 
of the thermal circle, but the dynamics in the intermediate regime is dominated by 
genus zero condensates that can materialize at the critical point. 

Around the critical point $R=R_c$, additional thermal states characterized by both non-trivial winding and momentum charges and masses
\be
m^2={1\over 4R_c^2}\left({R_c \over  R}-{R \over R_c}\right)^2,
\ee
 become light, giving rise to an enhanced $SU(2)_L$ gauge symmetry.
Thus the intermediate regime admits a well-defined description in terms of an $[SU(2)_L]_{k=2}$ worldsheet CFT, and the effective thermal theory at the critical point is in terms of an $SU(2)$ gauge theory with massless scalar fields in the adjoint representation. Each asymptotic thermal phase arises as a spontaneously broken symmetry phase as we deform away from the fermionic extended symmetry point. 

Depending on the number of non-compact directions, the extra massless thermal states induce a conical structure in the one-loop partition function or for one of its derivatives as a function of $R$ -- with three non-compact spatial directions the partition function is smooth but the fourth derivative of $Z$ is discontinuous (as a function of $R$) -- indicating a non-trivial phase transition between the two asymptotically cold regimes as the temperature varies. It turns out that the stringy transition at $R_c$ can be resolved in the presence of genus-0 condensates associated with the extra massless states, which mediate transitions between purely thermal winding and purely thermal momentum states. In \cite{KPT,KPT2,FKPT,CorfuLectures}, it was shown that the equations of motion at the critical point allow such condensates to be realized via non-zero spatial gradients of the corresponding extra massless thermal fields, leading to additional negative pressure contributions in the low energy effective action. 

In the Lorentzian description, where the temperature becomes time dependent,
these contributions can give rise to a spacelike brane (S-brane) configuration, materializing along a time slice at which the temperature reaches its critical value $\tilde T_c$ \cite{KPT,KPT2,FKPT,CorfuLectures}. 
The brane has finite but string-scale thickness in time, and hence in the low energy effective description, we can treat it as a $\delta$-function source (since resolving it would require probes with energies that approach the cut-off of the effective theory). Since the brane is spacelike, its associated energy density will vanish, but it will have negative pressure in the spatial dimensions. Away from the $\delta$-function source, the effective action is that of the usual low energy degrees of freedom of string theory i.e. of dilaton-gravity coupled to a thermal gas of pressure $\tilde p$ and energy density $\tilde \rho$. The S-brane can be viewed as the spacetime ``defect" associated with the transition between different geometrical phases of the underlying worldsheet CFT. In the low energy effective theory, it evidently mediates a transition from contraction to expansion by providing precisely the distributional surface energy momentum tensor required to satisfy the Israel junction conditions \cite{Israel}.  

In the adiabatic approximation, cosmological dynamics very different from the usual point particle physics induced dynamics can be obtained. One possibility is a solution in which the underlying thermal modulus $R$ runs from $0$ to $\infty$ \cite{KPT,KPT2, FKPT, CorfuLectures}. Correspondingly the (string frame) temperature grows from small values, reaching its maximal value, $\tilde T_c$, and then drops again to zero. At sub-critical values of $R$, the Universe is in a contracting phase, while for $R>R_c$ the Universe is expanding. At the critical value of $R$, the S-brane configuration provides the violation of the NEC which induces a transition from contraction to expansion (when the cosmology is viewed in the Einstein frame). The dilaton grows in the contracting phase and undergoes an elastic bounce across the brane. In the expanding phase,
it decreases and asymptotes to a constant value in the future. Hence the dilaton attains its maximal value at the transition point, as set by the brane tension \cite{KPT,KPT2, FKPT,CorfuLectures}. If this critical value of 
the dilaton is sufficiently small the perturbative analysis can be justified.
Therefore, within the setup of \cite{KPT, KPT2,FKPT,CorfuLectures}, a bouncing cosmology connecting two asymptotically cold phases results.

In \cite{KPT2} another class of non-singular string cosmologies was obtained. By utilizing various fluxes in the effective gauged supergravity  description at the extended symmetry point, an ``emergent" phase with constant string frame temperature $\tilde T$, corresponding to $R=R_c$, and constant string frame scale factor $\tilde a$, but with growing dilaton can be supported. The string coupling is very weak in the far past, and so this brany phase can last for a long period of time. The regime terminates when the gauge condensates decay once the dilaton has reached a sufficiently large value, with the Universe entering in the phase $R>R_c$ described above. The transition surface can be again described by an S-brane configuration. In the Einstein frame, the emergent phase describes a contracting Universe with increasing temperature.
In this paper however, we will focus on the analysis of the cosmological fluctuations in the first cosmological scenario described above, deferring the analysis for the second interesting possibility for future work.

\subsection{Effective action gluing the dual regimes via the S-brane }

We take note of the fact that the violation of the NEC responsible for the bounce follows from the existence of the S-brane sources localized in time, associated to non-trivial spatial gradients for the extra massless thermal scalars \cite{KPT,KPT2,FKPT,CorfuLectures}. This phenomenon is independent of the precise equation of state of the thermal system in the neighborhood of the critical point. Therefore, in order to be quantitative in what follows, we shall approximate the thermal system with that of massless thermal radiation up to the critical point\footnote{Justifiable in the low energy approximation, as any induced corrections to the effective action will be in the form of higher derivative terms suppressed by appropriate powers of the cutoff, set by the string scale.}. 

Since we are at weak string coupling and since the size of the spatial torus is much larger than the string scale, it is possible to derive an effective action for the cosmological dynamics which is valid both in the asymptotic radiation eras and also close to the critical temperature. In the string frame and in the thin S-brane approximation, this action takes the form \cite{KPT,KPT2,FKPT,CorfuLectures}:
\eq{sfa}{S = \int d^4x\sqrt{-\tilde g} e^{-2\phi}\Bigl[\frac{\tilde R}{2} +2\tilde\nabla_\mu\phi\tilde\nabla^\mu\phi\Bigr] + 
\int d^4x\sqrt{-\tilde g}\, \tilde p - \int d\tilde\beta d^3\xi \sqrt{\tilde \gamma}~e^{-2\phi}\, \kappa\, \delta(\tilde \beta- \tilde \beta_c),}
where $\tilde g$ is the determinant of the string frame metric; $\tilde R$ denotes the corresponding Ricci scalar and $\phi$ is the dilaton. 
The second term in the action represents the contribution of the thermal string gas. In terms of the inverse temperature, the pressure is given by
\be
\label{P}
\tilde p= {\Lambda \over \tilde \beta^4}
\ee
with $\Lambda=n^*\sigma_r$. 
The third term is due to the S-brane, which is localized at the spacetime hypersurface on which the $\sigma$-model temperature is uniform, equal to its maximal (critical) value: 
$\tilde \beta = \tilde \beta_c$. The coordinates
on the S-brane surface are denoted by $\xi^i$; $\tilde \gamma$ is the determinant of the induced metric
\eq{hdef}{\tilde \gamma_{ab} = \tilde g_{\mu\nu}\frac{\partial X^\mu}{\partial \xi^a}\frac{\partial X^\nu}{\partial \xi^b}\, ,}
with $X^{\mu}(\xi)$ being the embedding fields and $\kappa$  the brane tension. Further on we present a more convenient expression of the brane action by utilizing the thermal scalar potential $\psi$ \cite{Mukhanovbook,Dubovsky:2005xd,fv}, which will be useful for the study of cosmological fluctuations: 
\be
\tilde T^2= -\tilde g^{\mu\nu}\partial_{\mu} \psi \partial_{\nu}\psi. 
\ee
The details of this potential formulation are covered in Appendix \ref{A}. 

Based on the action above, we can obtain non-singular, homogeneous and isotropic bouncing solutions \cite{KPT,KPT2,FKPT,CorfuLectures}, whose corresponding perturbations are to be analyzed in this work. Working in conformal gauge, the background string frame metric takes the form   
\be
d\tilde s_0^2 \, = \, \tilde a^2(\tau) \bigl[ - d\tau^2 + dx^2\bigr] \, ,
\ee
where $\tilde a$ is the scale factor and $\tau$ is conformal time.
Without loss of generality, we can in this case identify the S-brane surface with the $\tau=0$ slice, at which the temperature reaches its critical maximal value, $\tilde T_c$, and in addition its first time derivative vanishes. The background solutions (denoted by the subscript zero) for the temperature, scale factor and dilaton are given by \cite{KPT,KPT2,FKPT,CorfuLectures}:    
$$
\ln{\left(\tilde T_c \over \tilde T_0\right)} =\ln{\left(\tilde a \over \tilde a_c\right)} ={ 1\over 2}~\left[ {\eta_+} \ln \left(1+ {|\tau| \over \xi_+} \right) -{\eta_-}  \ln \left(1+{|\tau|\over \xi_-}  \right) \right]
$$
\be
\label{background}
 \phi_0=\phi_c+{\sqrt{3}\over 2}~\left[ \ln \left(1+ {|\tau|\over \xi_+} \right) - \ln \left(1+ {|\tau|\over \xi_-} \right)  \right] \, ,
\ee
where the time scale $\xi$ of the problem is set by the inverse of the brane tension $\kappa$,
\be 
\xi={2\sqrt{3}\over \kappa \tilde a_c}\,,\qquad \xi_\pm=\xi\eta_\pm \, ,\qquad \eta_\pm=\sqrt{3}\pm 1.
\ee
The tension, the critical temperature and the maximal value of the dilaton $\phi_c$ are related as follows \cite{KPT}
\be
\kappa = 2\sqrt{6\Lambda}~\tilde T_c^{2}~e^{\phi_c}\, .
\ee

In the neighborhood of the brane $|\tau|\ll \xi$,
the string frame metric is regular while the dilaton exhibits (in the thin-brane approximation) a conical singularity:
$$
\ln{\left(\tilde a \over \tilde a_c\right)} = {1\over 4}\,{\tau^2\over \xi^2}+{\cal O}\left({|\tau|^3\over \xi^3}\right) 
$$
\be
 \phi_0 =\phi_c- {\sqrt{3}\over 2}\, {|\tau|\over \xi}+{\cal O}\left({\tau^2\over \xi^2}\right) \, .
\ee
The scale factor takes its minimal value $\tilde a_c$ at the brane. Far from the brane $|\tau|\gg \xi$,
the dilaton asymptotes to a constant, the temperature drops and the scale factor tends to infinity.

The thermal entropy in any co-moving cell of unit coordinate volume (physical volume $\tilde a^{d-1}$), 
$S_0=4\Lambda (\tilde a \tilde T_0)^{3}$, remains constant, also
across the bounce, and hence the minimal value of the scale factor satisfies
\be
\label{S0}
\tilde a_c \tilde T_c = {\left(S_0 \over 4 \Lambda \right)}^{1/3}\, .
\ee
In particular $\tilde a_c$ can be kept large in string units for large entropy $S_0$, which is taken to be one of the integration constants of the problem. The Ricci scalar (in string units) attains its maximal value at the brane 
\be
{\tilde R}=\kappa^2/4 = {\cal O}(g_c^2);  
\ee
the string frame metric exhibits no essential singularity.
{\it Both $g_s$ and $\alpha^{\prime}$ corrections remain under control,} provided that the critical value of the string coupling $g_c$ is sufficiently small.

To pass to the Einstein frame, we perform a conformal rescaling of the metric via the dilaton field as follows
\eq{conf}{g_{\mu\nu} = e^{-2\phi} \tilde g_{\mu\nu}.}
As a result, the Einstein frame inverse temperature $\beta_0$ and scale factor $a$ are dressed by the string coupling,
\eq{bet}{\beta_0 = e^{-\phi}\tilde \beta_0,~~~ a=e^{-\phi}\tilde a \, ,}
and so they develop conical singularities, as inherited from the dilaton profile. The conical singularities are resolved by the brane at $\tilde T_c$, which supports the extra massless thermal states.

\subsection{Isothermal S-brane action}

Before analyzing the fluctuations around the background solution, 
we must elaborate upon the precise form of the S-brane action. In Appendix \ref{A2} we derive the isothermal S-brane action in terms of the thermal potential $\psi$: 
$\tilde T^2 = -\tilde g^{\mu\nu}\partial_{\mu}\psi\partial_{\nu}\psi$ (equivalently, in the Einstein frame
$T^2 = -g^{\mu\nu}\partial_{\mu}\psi\partial_{\nu}\psi$). 
Its form is very elegant and suggestive:
\begin{eqnarray}
S_B&\!\!=\!\!&-\int d\tau d^3x \, \kappa \, e^{-2\phi} \sqrt{-{\rm det}(\tilde g_{\mu\nu})}~\left(\sqrt{-\partial_{\mu} \psi\partial^{\mu} \psi}\right) \delta(\psi-\psi_c)\nonumber \\
 &\!\!=\!\!&-\int d\tau d^3x \, \kappa \, e^{-2\phi} \sqrt{-\tilde g}~\tilde T \delta(\psi-\psi_c)\, , 
\end{eqnarray}
where $\psi_c\equiv \psi(\pi(\vec x),\vec x)$ involves the time  location $\pi(\vec x)$ of the brane. In the Einstein frame, the S-brane action takes the form
\be
S_B=-\int d\tau d^3x  \sqrt{-g}\, \kappa\, e^{\phi}\, T\, \delta(\psi-\psi_c). 
\ee
It is interesting to note that this action is manifestly diffeomorphism invariant in the four-dimensional sense, as well as being invariant under the rescaling of the thermal potential $\psi \rightarrow \lambda \psi$ -- the subsequent rescaling of the temperature, $T \rightarrow \lambda $$ T$, cancelling in the rescaling of the $\delta$-function.  

The contribution of the brane to the Einstein frame energy momentum tensor is given by
\be
[T^{\mu\nu}]_B={2 \over \sqrt{-g}}{\delta S_B \over \delta g_{\mu\nu}}= -\kappa e^{\phi}\left( g^{\mu\nu}+{1 \over T^2}\, 
 g^{\mu\sigma} g^{\nu \omega}\partial_\sigma\psi\partial_\omega\psi\right)\,T\,\delta(\psi-\psi_c).
\ee
From this expression, we conclude that the brane stress tensor takes the form of a surface ``perfect fluid''
\be
[T^{\mu\nu}]_B= p_B g^{\mu\nu}+( p_B + \rho_B)u^{\mu}u^{\nu},
\ee 
with four-velocity given by
\be
u^{\mu}={\partial^{\mu}\psi \over \sqrt{-\partial_{\nu}\psi\partial^{\nu}\psi}}\, ,
\ee
as in the case of the bulk thermal string fluid. More importantly, the brane gives rise to {\it non-trivial, localized negative pressure with vanishing zero energy density} 
\be
p_B=-\kappa e^{\phi} ~{T}~ \delta (\psi-\psi_c),~~~{  \rho_B}=0 ~~~~~~\left(\tilde p_B=-\kappa e^{-2\phi} ~{\tilde T}~ \delta (\psi-\psi_c),~~~{\tilde \rho_B}=0\right),
\ee
even in the presence of fluctuations. Contributions to the entropy conservation law localized at the brane vanish once the first time derivative of the dilaton reflects across the brane, as we show in  Appendix \ref{A4}. 

Converting the $\delta$-function constraint on $\psi$ into a $\delta$-function constraint on $\tau$,
$$
\delta(\psi-\psi_c) = { 1\over \abs \psi^{\prime }\abs}
\delta (\tau-\pi)~={  \sqrt{\abs \tilde g^{00}\abs} \over \tilde T_c} \delta (\tau-\pi),
$$
where $\pi(\vec x)$ is the location of the brane at $\vec x$ \footnote{We drop here quadratic terms in the spatial gradients of $\psi$ which do not contribute to the equations of motion for the metric and dilaton fluctuations at the linearized level.}, we can recast the brane action in the string frame in the following form:  
\be
S_B=- \int d\tau d^3x 
 \sqrt{\tilde g_3}\, \kappa\, e^{-2\phi}\,  {\tilde T\over \tilde T_c}~\delta(\tau-\pi)\, ,
\ee
where $\tilde g_3$ is the determinant of the spatial metric $\tilde g_{ij}$.
In the Einstein frame, the action is given by 
\be
S_B=- \int d\tau d^3x \, 
 \sqrt{ g_3}\, \kappa\, e^{\phi}\, {T\over  T_c}~ \delta(\tau-\pi) \, .
\ee
We are now ready to consider perturbations of the coupled string fluid/dilaton/S-brane action. 

\section{Perturbations}

The analysis for the evolution of the curvature perturbations is most usefully carried out in the Einstein frame, where the action takes the form 
\eq{bga}{S = \frac{1}{2}\int d^4x \sqrt{-g}\, \Bigl[R - 2\nabla_\mu\phi\nabla^\mu\phi\Bigr] 
+ \int d^4x \sqrt{-g}{\Lambda}{T^4} -\int d^4x  \sqrt{-g}\, \kappa \, e^{\phi}  T \, \delta(\psi-\psi_c) ,}
after we perform the conformal rescaling of the metric as in Eq. (\ref{conf}).
It is important to stress that in the above expression, the Einstein frame temperature 
$T$ is given in terms of the thermal potential $\psi$ as
\eq{tdefk2}{T ={\beta^{-1}} = \sqrt{-g^{\mu\nu}\partial_\mu\psi\partial_\nu\psi}.}
We are concerned here with linear cosmological perturbations around the homogeneous and isotropic background defined in Eq. (\ref{background}). 
Our system has two matter components -- the thermal string fluid characterized by its temperature and the dilaton field. Therefore we perturb the action to second order in the fluctuations around the background solution as:
\begin{eqnarray}
\label{perts}
g_{\mu\nu} &=& g^{(0)}_{\mu\nu} + h_{\mu\nu}\\
\beta &=& \beta_0 + \delta\beta \\
\phi &=& \phi_0 + \varphi \, .
\end{eqnarray}
For a system with no anisotropic stress (as ours), it is convenient to parametrize the metric fluctuations by introducing the gravitational potential 
$\Phi(x)$ \cite{Mukhanovbook,MFB,RHBrev2}:
\eq{lp}{ds^2 \, = \, -a^2(1+2\Phi)d\tau^2 + a^2(1- 2\Phi)dx^2 \, .}
For a discussion of the derivation in different gauges, we refer the reader to Appendix \ref{adm4}. The dilaton fluctuations are parametrized by $\p (x)$, while fluctuations of the temperature potential by $\sigma (x)$:
\eq{tdeltapsi}{\psi=\psi_0+\sigma .}
By utilizing the equations of motion of the metric $g_{\mu\nu}$, the dilaton field $\phi$ and the thermal potential $\psi$, 
it is straightforward to derive the equations 
of motion that determine the evolution of the fluctuations around the background, away from the location of the S-brane: $\tilde \beta \ne \tilde \beta_c$. Throughout prime superscripts will denote time derivatives and ${\cal H}=a'/a$. 
The linearized equations of motion for the fluctuations are given as follows.
\vskip .2cm
\noindent {\it Metric fluctuations:}
\eq{00}{\nabla^2\Phi - 3\mathcal H \Phi' - 3\mathcal H^2 \Phi = -\frac{a^2}{2}\delta T^0_0}
\eq{01}{\partial_i(\Phi' + \mathcal H \Phi) = \frac{a^2}{2}\delta T^i_0}
\eq{11}{\Phi'' + 3\mathcal H\Phi' + \mathcal H^2\Phi + 2\mathcal H'\Phi = \frac{a^2}{2}\delta T^i_i,}
{\it Dilaton fluctuations:}
\eq{dpp}{{\p'' + 2\mathcal H \p' -\nabla^2\p -4\Phi'\phi_0' = 0},}
 {\it Temperature fluctuations:}
\eq{vpp2}{{\sigma'' -\frac{1}{3}\nabla^2\sigma + 2\frac{\psi_0''}{\psi_0'}\sigma' - 2\Phi'\psi_0' - 6\Phi\psi_0''= 0}.}
The perturbed components of the energy momentum stress tensor are given in terms of $\Phi, \, \varphi$ and $\sigma$ by:
\eq{em00}{\delta T^0_0 = -\delta\rho = 2\Phi\frac{\phi_0'^2}{a^2} - 2\varphi'\frac{\phi_0'}{a^2} + 12\frac{\Lambda}{\beta_0^4}\frac{\delta\beta}{\beta_0}} 
\eq{emi0}{\delta T^i_0 = 2\partial_i\varphi \frac{\phi_0'}{a^2} + (\rho + p)\frac{\partial_i\sigma}{\psi_0'}}
\eq{emii}{\delta T^i_i = \delta p = -2\Phi\frac{\phi_0'^2}{a^2} +2 \varphi'\frac{\phi_0'}{a^2} - 4\frac{\Lambda}
{\beta_0^4}\frac{\delta\beta}{\beta_0} }
\eq{emij}{\delta T^i_{j\neq i} = 0.}

Using the constraint equation relating the temperature in terms of the thermal potential $\psi$, Eq. (\ref{tdefk2}), 
we can express the temperature fluctuations $\delta\beta$ in terms of $\Phi$ and $\sigma$
\eq{tprop}{{\frac{\delta\beta}{\beta_0} = - \frac{\delta T}{T_0} = \Phi - \frac{\sigma'}{\psi_0'}}.} 
The background thermal potential is given in terms of the background inverse temperature and scale factor by
\eq{b0ans}{\psi_0' = \frac{a}{\beta_0}= \left({S_0 \over 4\Lambda}\right)^{1\over 3},}
where $S_0$ is the background thermal entropy. Since the latter is conserved, $\psi_0'' = 0$. 
The entropy conservation law is nothing but the equation of motion for the thermal potential $\psi$. 
Since $\psi_0'' = 0$, the temperature fluctuations propagate according to the relativistic diffusion equation (see Eq. (\ref{vpp2})),
\eq{vpp3}{\sigma'' -\frac{1}{3}\nabla^2\sigma  - 2\Phi'\psi_0' = 0 \,.}

Considering only the very long wavelength perturbations, we may consistently neglect spatial derivatives in the equations of motion. 
In this long wavelength limit, the equations of motion simplify, yielding simple analytic expressions for the fluctuations. 
Indeed, taking into account  
that the background solution for the dilaton satisfies 
$\phi_0'' + 2\mathcal H \phi_0' = 0$, it follows that 
$\phi^{\prime}_0\propto 1/a^2$. Together with the fact that $\psi_0$ is constant, this implies that the long wavelength perturbations 
satisfy the following first order equations:
\eq{solsig}{\frac{\sigma'}{\psi_0'} = 2\Phi + c_\sigma ~~~~{\rm i.e.} ~~~~{{\frac{\delta\beta}{\beta_0} =- \Phi -  c_\sigma}}, }
\eq{solpp}{\p' = \phi_0'( c_\p +4\Phi )  \, ,} 
where $c_\sigma$ and $c_\p$ are integration constants.

Both of these equations follow in a natural way from basic properties of the combined dilaton/thermal gas system.  
Despite the dilaton motion and the presence of metric fluctuations, the thermal entropy $S$ is conserved, 
and so
\be
\left({S\over 4\Lambda}\right)^{1 \over 3}= {a(1-\Phi) \over \beta}={\rm constant}.
\ee
Introducing the time independent fluctuation $c_\sigma$ via
\be
\label{entro}
\left({S \over S_0}\right)^{1 \over 3}=1+c_{\sigma}
\ee
(where $S_0=4\Lambda(a/\beta_0)^3$ is the background thermal entropy defined in Eq. (\ref{S0})),  
yields:
\eq{Entropy} {1-\Phi={\left(S \over S_0 \right)}^{1\over 3} {\beta \over \beta_0}\qquad\Longrightarrow\qquad {\delta \beta \over \beta_0}=-(\Phi+c_{\sigma}) \, .}
Likewise in the long wavelength limit, the full dilaton equation   
\eq{dilaton}{\phi^{\prime \prime}+ \left(3 {~a^{\prime} \over a} - { ~N^{\prime}  \over N}\right) \phi^{\prime}=0, }
where $A=a(1-\Phi)$ and $N=a(1+\Phi)$, is integrable, giving  
\eq{dilatonSolution}{\phi^{\prime}=C_0\, (1+c_{\p}) \,{ N \over A^3} =\phi_0^{\prime}\, (1+ c_{\p} +4\Phi) \qquad\Longrightarrow \qquad\p^{\prime }=\phi_0^{\prime}\,(c_{\p}+4\Phi)\, .}
Here $C_0$ is an integration constant associated with the dilaton background solution, $\phi_0^{\prime}=C_0/a^2$, 
while $c_{\p}$ is the corresponding time independent fluctuation. 

Combining the equations for $\Phi$ and $\p$ (in the long wavelength approximation), and thanks to the fact that the equation of state of the thermal 
fluid satisfies $\rho_r=3p_r$ {\it even in the presence of fluctuations,} we obtain the main equation of interest of this section concerning the metric
fluctuation $\Phi$:
\eq{bvprop}{{\Phi'' + 4\h\Phi' + 2(\h^2 + \h' - \phi_0'^2)\Phi = {2\over 3}\phi_0'^2 c_\p}.}
Denoting $\epsilon = \epsilon(\tau)$, the background solutions (\ref{background}) can be written as follows:
\begin{eqnarray}
\label{psol}
\phi_0^{\prime}(\tau) \!\!&=&\!\!-\epsilon\,{\sqrt{3}\,\xi \over y^2}= \frac{\sqrt{3}}{2}\left[\frac{1}{\tau +\epsilon\,\xi_+} - \frac{1}{\tau +\epsilon\, \xi_-} \right]\label{hc} \\ 
\h (\tau)\!\!&=&\!\!\epsilon\,{\sqrt{y^2+\xi^2}\over y^2} \label{hsol} =\frac{1}{2}\left[\frac{1}{\tau +\epsilon\, \xi_+} + \frac{1}{\tau +\epsilon\, \xi_-}
 \right]\\ 
 \h'(\tau)\!\! &=&\!\! -{y^2+2\xi^2 \over y^4}\label{hpsol} =-\frac{1}{2}\left[\frac{1}{(\tau +\epsilon\,\xi_+)^2} + \frac{1}{(\tau +\epsilon\, \xi_-)^2}\right]~,
\end{eqnarray}
where we have introduced the convenient variable $y$ whose relation to the background scale factor $a$ is given by   
\be
\label{y}
y^2=\big(\abs \tau\abs +\xi_+\big)\big(\abs\tau\abs+\xi_-\big),~~~~~~y={y_c\over a_c}\, a, ~~~~~y_c^2=\xi_+\xi_-=2\xi^2~.
\ee
Eq. (\ref{bvprop}) then becomes
\eq{peom}{\Phi'' + 2\Bigl(\frac{1}{\tau +\epsilon \xi_+} + \frac{1}{\tau +\epsilon\xi_-}\Bigr)\Phi' - 2\Bigl(\frac{1}{\tau +\epsilon\xi_+} - \frac{1}{\tau +\epsilon\xi_-}\Bigr)^2\left(\Phi+{c_\varphi\over 4}\right) = 0\, ,}
which together with Eq. (\ref{solpp}) for $\p$ are solved analytically in Appendix \ref{B}. The solutions can be expressed in terms of 
three integration constants $C_{1,2,3}$:
\begin{eqnarray}
\label{solphi}\Phi \!\!&=&\!\!C_1 \, \xi^2 \phi_0^{\prime} \mathcal H + C_2 \, (1+ 4\xi^2 \mathcal H^{\prime}\,)-\frac{c_\p}{4}\,,\\\nonumber\\
\label{solp}\p\!\!&=&\!\!-  C_1 \,\xi^2{\phi_0^{\prime}}^2-4C_2\,\phi_0 ^{\prime}\mathcal H (y^2-2\xi^2)+C_3\,.
\end{eqnarray}

The coefficient $C_2$ is not independent of the other integration constants, as follows from the Friedmann equation (\ref{00}) (and (\ref{em00})). 
The latter is a first order inhomogeneous equation, whose homogeneous part admits  $\Phi_1=C_1 \, \xi^2 \phi_0^{\prime} \mathcal H$ as a general solution. 
What remains is to find $C_2$ such that 
\be
\Phi_{\rm part}= \Phi_2 -{c_\varphi\over 4}=C_2 (1+ 4\xi^2 \mathcal H^{\prime}\,)-{c_\varphi\over 4}
\ee
is a particular solution of the full equation. This constraint yields 
\be
C_2= {c_{\p}\over4}-{2\over 3}c_{\sigma} \, .
\ee 

In the expressions (\ref{solphi}), (\ref{solp}) and (\ref{solsig}) for $\Phi$, $\p$ and $\delta\beta$, which are valid in the long wavelength approximation, it is understood that the integration constants are space-dependent functions. Moreover, two sets of integration constants must be introduced, $\{c_\sigma^-, c_\varphi^-, C_1^-, C_3^-\}$ and $\{c_\sigma^+, c_\varphi^+, C_1^+, C_3^+\}$, in order to describe the fluctuations before and after the S-brane, {\em i.e.} for $\tau<\pi$ and $\tau>\pi$. It turns out that the S-brane imposes well defined relations among the two sets. These relations turn out to be fundamental since they fix the spectrum of fluctuations in the expanding phase in terms of the primordial fluctuations in the contacting phase. 

\section{Matching the Fluctuations}

We are obliged to match the metric perturbations across the S-brane consistently with the Israel junction conditions \cite{Israel}. 
This implies \cite{HV, DM} that for the full solution, 
the induced metric $\gamma_{ab}$ on the hypersurface defining the locus of points where the string frame temperature reaches its critical value, 
$\tilde T= \tilde T_c$, must be continuous across the bounce. The extrinsic curvature must jump according to a surface tensor.
 The Israel junction conditions precisely enforce Einstein's equations 
on the gluing hypersurface, and require a surface energy momentum tensor to compensate for the distributional jump in the Einstein tensor. In our model, 
the S-brane provides us exactly with such a surface energy momentum tensor\footnote{Specifically, the S-brane mediates the contracting and expanding phases 
with an instantaneous violation of the NEC. Because this violation occurs over a time scale that approaches the cut-off of the low energy effective theory, 
dangerous instabilities do not have time to develop and the system is evidently stable, as one would expect given the underlying string theoretic consistency of the model.}. 
Having satisfied the necessary junction conditions for the background, we must do so for the perturbations. 

\subsection{The equations of motion in the string frame}

 The junction conditions 
are more transparent in the string frame. As we are going to see, the first time derivative of the dilaton jumps, 
while the string frame metric and its first time derivative are regular. 
These follow from the gravitational and dilaton equations of motion in the string frame. The metric in this frame (including the perturbations) 
takes the following diagonal form
\be
d\tilde s^2=-\tilde N(\tau,\vec x)^2d\tau^2+\tilde A(\tau,\vec x)^2dx^2\, .
\ee
Using the fact that the equation of state for the thermal bulk fluid is $\tilde \rho_r$=$3\tilde p_r$,  
we obtain the following equations in the long wavelength approximation.

$~$\\
 {\it $\tilde N$-equation:}
\be
\label{SFFried}
3 \left({\tilde A^{\prime }\over \tilde A}\right)^2=6{\, \tilde A^{\prime }\over \tilde A}\phi^{\prime}-2{\phi^{\prime}}^2+e^{2\phi}\tilde N^2 \tilde\rho_r\,.
\ee
$~$\\
 {\it The trace equation (modulo the dilaton equation):}
\be
{\, \tilde A^{\prime \prime}\over \tilde A}+ {\, \tilde A^{\prime }\over \tilde A}\left( {\, \tilde A^{\prime }\over \tilde A} -{\, \tilde N^{\prime}\over \tilde N}\right) = {2\over 3}{\phi^{\prime}}^2\,.
\ee
$~$\\ 
{\it The $\phi$-equation (modulo the trace equation):}
\be
\phi^{\prime \prime}+ \phi^{\prime} \left(3 {\, \tilde A^{\prime }\over \tilde A}-  { \,\tilde N^{\prime}\over \tilde N}  -2 \phi^{\prime}\right)= -\frac{\kappa}{2}\, \tilde N\,{\tilde T\over \tilde T_c}\delta (\tau-\pi)\,.
\ee

The dilaton equation in particular shows that the first time derivative 
of the dilaton is discontinuous across the transition surface. 
The discontinuity is resolved by the presence of the spacelike brane whose locus $\tau=\pi(\vec x)$ 
is defined via the equation $\tilde \beta(\pi,\vec x)=\tilde \beta_c$. The jump in the
time derivative of the dilaton is determined by the tension of the brane:
\be
\label{kphi'}
{\phi^{\prime}(\pi_{-},\vec x)-\phi^{\prime}(\pi_{+},\vec x)\over \tilde N(\pi,\vec x)}={\kappa(\vec x)\over 2}\, .
\ee

In the remainder of this section, all $\vec x$-dependences are dropped for notational convenience. The expressions of $\tilde A$ , $\tilde N$  and $\tilde \beta$ in terms of $\tilde a$, $\tilde \beta_0$ and the fluctuations $\Phi$, $\varphi$ are given by
\begin{eqnarray}
\tilde A\!\!&=&\!\!e^{\phi}\, A=e^{\phi}\, a\, (1-\Phi)=\tilde a\, (1-\Phi+\p)\,,\\\nonumber\\
\tilde N\!\!&=&\!\!e^{\phi}\, N=e^{\phi}\, a\, (1+\Phi)=\tilde a \, (1+\Phi+\p)\,,\\\nonumber\\
\tilde \beta \!\!&=&\!\!e^{\phi}\, \beta=e^{\phi}\, (\beta_0+\delta\beta_0)=\tilde\beta_0\, (1-c_{\sigma}-\Phi+\p )\,.
\end{eqnarray}
The second time derivative of $\tilde A$ appears in the trace equation. 
Since there is no localized contribution from the brane in this equation, $\tilde A$ and its first time derivative
must be continuous across the brane. Because $\tilde a$ and its first time derivative are also continuous, 
the quantity $\Phi-\p$ must be regular as well. 

The stringy isothermal constraint turns out to be crucial. The main observation is that the conservation of the background thermal entropy, 
as well as the conservation of entropy and energy in the full system (including the fluctuations), strongly constrains the fluctuations. 
Recall that we introduced the constant (in time) fluctuation $c_{\sigma}$ to parametrize the difference between the background thermal entropy 
and the full thermal entropy, Eq. (\ref{entro}).
Since both $S_0$ and $S$ are conserved,
we obtain the following relation between the full scale factor $\tilde A$ and $\tilde a$:
\be
\tilde A(\tau_1) \, \tilde T(\tau_1)=\tilde a(\tau_2)\,  \tilde T_0(\tau_2)\,(1+c_{\sigma}), ~~~{\rm for~any}~ \tau_1~ {\rm and} ~ \tau_2\, .
\ee
This relation provides an important constraint, which is {\em not present} in other brane mediated bouncing models studied previously in the literature. Indeed, choosing $\tau_1$ to coincide with the locus of the perturbed S-brane ($\tau_1=\pi$) and $\tau_2=0$ (the S-brane locus in the absence of fluctuations), 
we get
\be
\label{pio}
\tilde A(\pi) \, \tilde T(\pi)=\tilde a(0)\,  \tilde T_0(0)\,(1+c_{\sigma})\, .
\ee
What is particular for the isothermal stringy S-brane under consideration follows from the existence of a maximal, critical temperature $\tilde T_c$, which is 
attained both at the locus of the perturbed S-brane as well as at the locus of the background brane: $\tilde T({\pi})=\tilde T_0(0)=\tilde T_c$. Eq. (\ref{pio}) then implies 
\be
\label{entropyCons}
\tilde A(\pi)=\tilde a(0)\, (1+c_{\sigma})\, .
\ee
On the other hand, up to the linear order in the fluctuations, $\tilde A(\pi)$ is given by
\be
\tilde A(\pi)=\big[\, \tilde a(0)+\tilde a^{\prime}(0)\pi \, \big](1-\Phi (0)+\p(0))=\tilde a(0)(1-\Phi (0)+\p(0))\, ,
\ee
where we used the property of the background $\tilde a^{\prime}(0)=0$. Therefore, we obtain the relation $c_{\sigma}=-\Phi(0)+\p(0)$, which in turn yields
\be
\label{C3}
C_3=3c_{\sigma}-c_{\varphi} \, .
\ee
 
It will be useful for later considerations to translate Eq. (\ref{entropyCons}) in the Einstein frame. To this end, we introduce the coefficient $C_A$:
$$
A(\pi)=\tilde A(\pi)e^{-\phi(\pi)}=\tilde a_c(1+c_{\sigma})e^{-\phi(\pi)}=a_c(1+c_{\sigma})e^{-\phi(\pi)+\phi_c} \equiv a_c(1+C_A)\, ,
$$
i.e.
\be
A(\pi)= a_c(1+C_A)\qquad{\rm and}\qquad e^{\phi(\pi)-\phi_c}={1+c_{\sigma} \over 1+C_A}.
\ee
Another useful quantity, related to the locus $\pi$ of the S-brane, is the coefficient $C_{a}$ defined via
\be
\label{Ca}
a(\pi)=a(0)(1+C_{a})\qquad {\rm i.e.}\qquad y(\pi)=y_c(1+C_{a})\, .
\ee
The explicit values for $C_A$ and $C_{a}$ in terms of the other fluctuations and  $\pi$ are determined in the following section.

\subsection{Junction conditions at the S-brane }
As we already stated the junction conditions for the metric and the dilaton field are more transparent in the string frame. The relevant quantities are the three dimensional spatial metric with scale factor $\tilde A$ and the dilaton field $\phi$. 
We now display the junction conditions:

$~$\\
{($i$)  \it Continuity of the metric :}  $\tilde A(\pi_-)=\tilde A(\pi_+)$

$~$\\
{ ($ii$) \it   Continuity of the time derivative of the metric :} $\tilde A^{\prime}(\pi_-)=\tilde A^{\prime} (\pi_+)$

$~$\\
{($iii$)   \it  Continuity of $\phi$ :}  $\phi(\pi_-)=\phi(\pi_+)$

$~$\\
{ ($iv$)  \it   Discontinuity condition for $\phi^{\prime}$ :}  
$\left[   \phi^{\prime}(\pi_-)-\phi^{\prime}(\pi_+)\right] /\tilde N(\pi)={\kappa / 2}$.
\\

The following comments are in order.

$~$\\
$\bullet$ Condition ($i$) and Eq. (\ref{entropyCons}) imply that  
\be
\tilde A(\pi_-)=\tilde A(\pi_+)=\tilde a(0)(1+c_{\sigma}^\pm)\qquad \mbox{\rm i.e.}\qquad c_{\sigma}^-=c_{\sigma}^+:=c_{\sigma} \, .
\ee
This result is very important since it shows that there is no entropy production by the S-brane.

$~$\\
$\bullet$ Condition ($ii$) together with the fact that $\tilde T(\pi_{\pm})=\tilde T_c$, and also the fact that $\tilde A$ 
must take its minimal value {\it without any discontinuity} in its first time derivative (as follows from the trace equation), 
imply a much stronger condition:
\be
\label{aprime0}
\tilde A^{\prime}(\pi_-)=0~~~~~{\rm  and} ~~~~~\tilde A^{\prime} (\pi_+)=0\, .
\ee
As we will see the constraints above determine $C_a$ in terms of the other fluctuations. 
 
$~$\\
$\bullet$ The continuity of $\phi$ at the locus of the brane fixes the maximal value of the string coupling in terms of the fluctuations. 

 $~$\\
$\bullet$  Finally the discontinuity condition for the first time derivative of the dilaton is the only non trivial junction condition at the S-brane.
\\

First we show that $C_a$ is determined by the requirement that ${\tilde A^{\prime}}(\pi)=0$, from which it follows that
\be
{\tilde A^{\prime} \over \tilde A}\Big\abs_{\pi}= \left[\mathcal H(\pi)+\phi^{\prime}_0(\pi)-\Phi^{\prime}(0)+\p^{\prime}(0)\right]=0\, .
\ee
Observe that the term $\mathcal H+\phi^{\prime}_0={\tilde a^{\prime}/ \tilde a}$ at the locus of the S-brane is given in terms of $C_a$ by the expression
\be
\mathcal H(\pi)+\phi^{\prime}_0(\pi)={\epsilon(\pi)\left(\sqrt{(y_c^2(1+2C_a)+\xi^2)}-\sqrt{3}\,\xi \right) \over (1+2C_a)\, y_c^2}\,.
\ee
Since $y^2_c=2\xi^2$,
\be
\mathcal H(\pi)+\phi^{\prime}_0(\pi)={ \epsilon(\pi)\xi\left(\sqrt{3+4C_a}-\sqrt{3}\, \right) \over  2\xi^2(1+2C_a)}=\epsilon(\pi){{\sqrt{3}\over 3\xi}}~C_a\,.
\ee
Using the first order equation for $\Phi$ 
$$
\Phi^{\prime}=-\mathcal H~{3y^2+4\xi^2\over y^2+\xi^2} \left(\Phi +{c_{\p}\over 4}\right)+\mathcal H{3C_2\, y^2\over \xi^2+y^2},
$$
and Eq. (\ref{solpp}), and taking into account the fact that at our level of approximation we may set in these equations $\mathcal H=-\phi_0^{\prime}$, we obtain
\be
-\Phi^{\prime}(0)+\p^{\prime}(0)=-\epsilon(\pi){\sqrt{3}\over 3\xi }\big(\Phi_1(0)+\Phi_2(0)+3C_2\big) .
\ee
Combining the results above, we obtain the desired relation between $C_a,~ C_1$ and $C_2$:
\be
C_a=\Phi_1(0)+\Phi_2(0)+3C_2=-{3\over 4} C_1^\pm\, ,
\ee
and conclude that $C_1^-=C_1^+:= C_1$. 

The remaining constraint involves the first time derivative of $\phi$ at the S-brane:
\be
\label{dila}
\phi^{\prime}(\pi_{\pm})=\phi_0^{\prime} +\p^{\prime}\Big \abs_{\pi_{\pm}}=
\mp\sqrt{3}\, \xi(1+c_{\p} ) e^{2\phi}{\tilde N\over \tilde A^3}\Big \abs_{{\pi}}\,={\mp\sqrt{3}\, \xi \over y_c^2(1+2C_a)}\left(1+ 4(\Phi(0) +{c_{\p}\over 4})\right)\, .
\ee
Dividing by $\tilde N(\pi)$,
$$
\tilde N(\pi)= 
{\tilde a}_c \left(1+2\Phi(0) +c_{\sigma}\right)\, ,
$$
we obtain:
\be
{\phi^{\prime}\over \tilde N}\Big \abs_{{\pi_\pm}}= \mp \Big(1+3c_{\sigma}- c_{\p} \Big) {\kappa_c \over 4}\, , 
\ee
where we use the relation ${3C_2 -\frac{3}{4}c_{\p}+2c_{\sigma}=0}$, as imposed by the Friedmann equation.  
Here $\kappa_c$ is the value of the unperturbed brane tension. The tension of the perturbed brane $\kappa$ 
scales with the full string coupling $e^{\phi}$, as follows from the Friedmann equation (\ref{SFFried}) and the matching condition
(\ref{kphi'}). So $\kappa$ becomes:
\be
{\kappa\over 2}=
{\kappa_c \over 2}e^{\phi_0(\pi)-\phi_0(0)+\p(\pi)}~={\kappa_c\over 2}\,(1-C_a+\p(0))={\kappa_c\over 2}\,(1+C_3). 
\ee
As shown in Appendix \ref{A4}, there is no thermal entropy production when crossing the brane. Using the constraints Eq. (\ref{aprime0}), this translates into reflective boundary conditions for the dilaton slope, $\phi'(\pi_-)=-\phi'(\pi^+)$, thus imposing $c_\varphi^-=c_\varphi^+:= c_\varphi$ (see Eq. (\ref{reflect})). The relation (\ref{C3}) implies $C_3^-=C_3^+:=C_3$, consistently with the continuity condition ($iii$) of the dilaton. Therefore, the matching conditions impose $\{ c_{\sigma}^-,c_{\varphi}^-,C^-_{1,2,3}\}=\{ c_{\sigma}^+,c_{\varphi}^+,C^+_{1,2,3}\}$.

For academic purposes we verify the discontinuity conditions for $\phi'$ and $A'$ in the Einstein frame. Dividing Eq. (\ref{dila}) with $N(\pi)$,
$$
N(\pi)={ a}_c(1+C_A) (1+2\Phi(0)) = { a}_c(1+C_A+2\Phi(0)),
$$
we obtain:
\be
{\phi^{\prime}\over  N}\Big \abs_{{\pi_\pm}}=\mp \left(1+{ c_{\p}\over 2} -C_A+2\Phi_2(0)\right) \left({\tilde a_c \over a_c} \right){\kappa_c \over 4} ~=\mp \left(1+{ c_{\p}\over 2}-C_A+2\Phi_2(0) \right) { \kappa_c \,e^{\phi_c} \over 4}\, . 
\ee
In the Einstein frame, the S-brane contributions appear with an extra factor of the string coupling $e^{\phi(\pi)}$. 
Therefore, the relevant term involved in the matching condition is:
\be
{\kappa \over 2}~ [e^{\phi(\pi)}]={\kappa_c e^{\phi(\pi)-\phi_c}\over 2}~[e^{\phi(\pi)-\phi_c}]\,e^{\phi_c}= {\kappa_c \,e^{\phi_c}\over 2}
( 1+2C_3)\, ,
\ee
imposing
\be
{c_{\p}\over 2}-C_A+2\Phi_2(0) = 2C_3\quad \Longrightarrow \quad  -C_A+4c_{\sigma}-c_{\p}=2C_3. 
\ee
Given the definition of $C_A$, it follows that 
\be
1+C_A=(1+c_{\sigma})e^{-\phi_0(\pi)+\phi_0(0)-\p}= 1+c_{\sigma}-C_3\qquad \Longrightarrow\qquad C_A=c_{\sigma}-C_3 .
\ee
Using the value of $C_A$, we obtain precisely the  constraint already derived in the string frame. The matching conditions for $A$ and $A^{\prime}$ follow from those of the dilaton and $\tilde A$.

For completeness, let us verify that the matching conditions are compatible with the Israel junction conditions across
the brane hypersurface $\tau - \pi(\vec x)=0$, in the Einstein frame. The Israel junction conditions require that the induced metric on this
hypersurface is continuous across it and that the extrinsic curvature jumps according to a surface tensor. In longitudinal gauge
and at long wavelengths (where we drop spatial derivatives) these conditions become
\be
g_{ij}(\pi_+)=g_{ij}(\pi_-)
\ee 
and 
\be
K_{ij}(\pi_+)-K_{ij}(\pi_-)=\left({(A^2)^{\prime} \over  2N}\Big \abs_{{\pi_+}}-{(A^2)^{\prime} \over  2N}\Big \abs_{{\pi_-}}\right)\delta_{ij}=-{1 \over 2}[{\cal {T}}^B]_{ij},
\ee
where $[{\cal {T}}^B]_{ij}$ is the brane stress tensor (modulo the delta function):
\be
[{\cal {T}}^B]_{ij} = -\kappa e^{\phi} g_{ij}\Big \abs_{{\pi}}. 
\ee
The first condition is guaranteed by the continuity of the full Einstein frame scale factor $A$ across
the brane (which follows from the continuity of the dilaton and the string frame scale factor).
For the second condition, we use the relation $A=e^{-\phi}\tilde A$ and the matching conditions $\tilde A^{\prime}(\pi_+)=\tilde A^{\prime}(\pi_-)=0$ to
get that the LHS is given by
\be
\left(-{\phi^{\prime} A^2\over  N}\Big \abs_{{\pi_+}}+{\phi^{\prime}A^2 \over  N}\Big \abs_{{\pi_-}}\right)\delta_{ij}={\kappa e^{\phi} \over 2}g_{ij}\Big \abs_{{\pi}},
\ee
where we have used Eq. (\ref{kphi'}). Evidently the requirement on the extrinsic curvature is also satisfied. Since we verified the 
conditions for the full fields $A$ and $\phi$, the constraints on the perturbations are also compatible with the analysis above.

It is important to notice that in the long wavelength approximation $C_1$ is not restricted by any matching condition at the locus of the S-brane, 
nor by the Friedmann equation. Its role lies in the determination of $\pi$, otherwise it is arbitrary. 
Using the relation
\be
y^2(\pi)=(1+2C_a )y^2(0), ~~~~~\Big(\abs \tau\abs+\xi_+\Big)\Big(\abs \tau\abs+\xi_-\Big)\Big \abs_{\pi} =(1+2C_a)2\xi^2\, ,
\ee
we can derive $\pi$ in terms of $C_a$. At the linear level of the approximation we find ($C_a=-{3\over4} C_1$):
\be
\abs \pi \abs ={2 \,\xi \over \sqrt{3}}C_a =-{\sqrt{3}\,\xi \over 2}  C_1.
\ee
 
\subsection{Summary of the results}

We derived the metric and dilaton fluctuations in analytic form:  
$$
\Phi =C_1 \, \xi^2 \phi_0^{\prime} \mathcal H + C_2 (1+ 4\xi^2 \mathcal H^{\prime}\,) -\frac{c_\p}{4} \, ,
$$
\be
\p =- C_1 \,\xi^2{\phi_0^{\prime}}^2-4C_2\phi_0 ^{\prime} \mathcal H (y^2-2\xi^2)  +C_3 \, .
\ee
The coefficients $C_2$ and $C_3$ depend on the two independent parameters 
$c_{\sigma}$, $c_{\p}$, while $C_1$ determines the locus of the S-brane:
\be
\pi=-\epsilon(\pi){\sqrt{3}\, \xi \over 2} C_1,  ~~~~~~C_2= {c_{\p} \over 4}-{2c_{\sigma} \over 3},~~~~~~C_3=3c_{\sigma}-c_{\p}\, .
\ee

These relations strictly restrict the strength of the fluctuations. To see this, 
it is useful to express $\phi_0$ and $\mathcal H$ in terms of $y$ and $\xi$, with $\xi \tilde a_c$ setting the proper time-scale of the problem 
$$
{1\over \xi \tilde a_c}=\sqrt{2\Lambda} ~\tilde T_c^2 ~e^{\phi_c} \, .
$$
The maximal, critical temperature, $\tilde T_c$, is of the order of the string scale $M_{\rm str}$, while $\Lambda$ is proportional to the number of the massless degrees of freedom, 
which is typically more than $10^3$ in semi-realistic string models. Also  $e^{\phi_c}=g_c$ is the maximal value of the string coupling, which can be taken 
to be small 
(say in the range $10^{-4} < g_c < 10^{-2}$) in order to ensure the validity of string perturbation theory. 
The relation $(\tilde T_c \tilde a_c)^3=(\tilde T  \tilde a)^3$ defines the entropy of the cosmology, which is a conserved quantity. 
In any realistic model the entropy at late times must be enormous. 
Since there is no extra entropy production in our cosmological model, 
the entropy at $\tilde T_c$ must be very big. On the other hand the value of $\tilde T_c$ is fixed, and so it implies that $\tilde a_c$ (which is the minimal value for the scale factor) 
is very big in string units. 
These observations show that the intermediate regime which includes the S-brane is short in time: 
$\Delta t_{\rm proper}\sim 2\xi \tilde a_c$, which can be, depending on the
value of the critical coupling, as small as a few string lengths. 

Let us examine in more detail the structure of perturbations and their evolution, in terms of $C_1$, $c_{\p}$, $c_{\sigma}$, $y$ and $\xi$: 
\be \label{Phieq}
\Phi =-C_1\,{\xi^3 \sqrt{3(y^2+\xi^2)} \over  y^4} 
 +{(8 c_{\sigma}-3c_{\p})\over 3} \left({\xi^2 (y^2+2\xi^2)\over y^4}  \right)  -{2\over 3}c_{\sigma} ~,
\ee
\be \label{phieq}
\p =-C_1  {3\xi^4 \over y^4 } 
-{(8 c_{\sigma}-3c_{\p})\over 3}\left({\xi \sqrt{3(y^2+\xi^2)} (y^2-2\xi^2)\over y^4 }\right) +(3c_{\sigma}-c_{\varphi}).
\ee

$~$\\
Some comments are in order:

$~$\\
(i) In the above expressions the variable $y^2$ is bounded from below: $y^2>2\xi^2$. 
This implies that the dangerous, growing fluctuation modes proportional to $C_1$ and $C_2$ in the contracting phase are bounded. 
Their maximal values are reached at the S-brane regime, where $y^2\sim 2\xi^2$. 

$~$\\
(ii) The coefficients in front of the $C_2$-growing modes are not independent but are correlated with other parameters thanks to:
1) the thermal entropy conservation in the system and 2) the energy conservation of the system (dilaton plus thermal fluid).

$~$\\
(iii)
The continuity condition $\tilde A(\pi_+)=\tilde A(\pi_-)$ implies $c_{\sigma}^+=c_{\sigma}^-=c_{\sigma}$. The continuity condition for $\phi$ imposes $C_{3}^\pm=C_3$. 
The matching relation $c_\p^\pm=c_{\p}$ follows from the fact that there is no entropy production from the S-brane.  
Finally, the continuity of $\Phi$ imposes that $C_1^\pm=C_1$. Therefore the long wavelength fluctuations are mirror reflective across the S-brane.

$~$\\
(iv) The entropy and energy conservation laws lead to correlations among $C_2, C_3, c_{\p}$ and $c_{\sigma}$, 
in accordance with the maximal temperature bound (imposing also $y^2>2\xi^2$). 
Their explicit expressions show clearly that the growing modes proportional to $C_2$, and also $C_1$, never dominate the system. 
Their magnitudes are always sub-dominant for $y\gg\xi$. They become relevant around the S-brane regime, where their amplitudes reach those of the constant modes. 

$~$\\
(v) Although the $C_2$-growing modes are irrelevant away from the S-brane regime, 
their presence is important closed to the S-brane regime. 
Their contribution is essential for the realization of the matching conditions at the S-brane. 
These growing modes are mixed with the constant modes at the locus of the S-brane.
 
$~$\\
(vi) The growing modes associated to $C_1$ do not participate in the junction conditions at the S-brane. However,  
$C_1$ defines the time displacement of the S-brane.

$~$\\
(vii) The important output of our analysis is that the relevant primordial perturbations $\Phi$ and $\p$ in the contracting phase, 
which persist in the expanding phase after crossing the S-brane are:
 \be \label{latelimit}
\Phi_k=-{2\over 3}c_{\sigma}(k),~~~~~~\p_k=- c_{\p}(k)+3c_{\sigma}(k) \, .
\ee

\subsection{Curvature perturbation} 
\label{ZETA}
 
Following \cite{Mukhanovbook,MFB,RHBrev2}, we define the gauge invariant quantity 
 \be
 \zeta=\Phi + {2 \over 3}{\Phi'+\h \Phi \over \left(1 + {p\over \rho}\right)\h} =  \Phi+{\h\Phi^{\prime} +\h^2\Phi
\over \h^2-\h^{\prime}},
\ee
where $\rho$ and $p$ are the total energy density and pressure, which is equal to minus the curvature perturbation on
co-moving slices: $\zeta = -{\cal{R}}_c$. 
Using the latter expression, we can verify that in the long wavelength limit, where we neglect spatial gradients,
the $\Phi$-mode proportional to $C_1$ does not contribute to $\zeta$. 

We now proceed to compute $\zeta$ in the thermal string gas/dilaton phase of the string cosmology.
Using the Friedmann equation (\ref{00}), we obtain in the long wavelength limit 
\be
\zeta = \Phi -{\delta\rho \over 3(\rho+p)}.
\ee
Furthermore utilizing the expression for the total energy perturbation $\delta \rho$, Eq. (\ref{em00}), 
and equations $\p'=\phi_0'(c_\p +4\Phi)$, $\delta \beta=-\beta_0(\Phi + c_\sigma)$,
we get
\be
\zeta = { 1\over \rho+p}\left(-2c_\p {\phi_0'^2 \over a^2} - {12 \Lambda \over \beta_0^4}c_\sigma \right).
\ee
It is clear from this result that the continuity of the perturbations $c_\p$ and $c_\sigma$ ensures the continuity of the curvature perturbation across the S-brane\footnote{Note that $\rho + p$ (the total energy density of the string gas + dilaton fluid + its total pressure) is non-zero at either side of the brane, and $\zeta$ is finite on both sides of the S-brane.}. 

In order to isolate adiabatic perturbations from ``entropy'' perturbations, it is useful to parametrize $\zeta$ in terms of
the coefficient 
\be
C_e=3c_\sigma-c_\p
\ee
and $c_\sigma$. Notice that the matching conditions at the S-brane impose that $C_e$ is equal to the coefficient  
$C_3$, which sets the asymptotic value of the 
dilaton fluctuation $\p$ early on in the contracting thermal string gas/dilaton phase (and also in the expanding phase at latter times) --
see Eq. (\ref{C3}). This fact will have important
implications on realizing the matter bounce scenario in the following section.
In terms of $C_e$ and 
$c_\sigma$, $\zeta$ is given by
\be
\label{zeta}
\zeta = -c_\sigma + {2C_e\phi_0'^2\over 3a^2(\rho+p)}=-c_\sigma + {C_e \over 3+ 4\big({a \over a_c}\big)^2}=
-c_\sigma + {C_e \over 3+ 4\big({\tilde T_c \over \tilde T}\big)^2}.
\ee
The last expression is in terms of the ratio of the temperature to the critical one in string frame.

Thus when $3c_\sigma=c_\p$ ($C_e=0$), $\zeta=-c_\sigma$ is conserved on superhorizon scales, the characteristic behavior of 
the non-trivial adiabatic mode.
In this case $\zeta = \zeta_r$, where 
\be
\zeta_r=\Phi - {1 \over 3} {\delta \rho_r \over (\rho_r + p_r)},
\ee  
which as in the case of any isentropic fluid is a conserved quantity on superhorizon scales on its own.
For ``entropy'' modes on the other hand, $C_e\ne 0$. When $\tilde T \ll \tilde T_c$, these modes are suppressed by a factor $(\tilde T / \tilde T_c)^2$. 

Our conclusion is that the primordial curvature perturbation which persists in the expanding phase after the S-brane crossing is given by
\be
\zeta_k\simeq -c_\sigma(k)\,.
\ee
When $C_3=0$, the above relation becomes an exact equality at long wavelengths. As we will see in the next section, this curvature perturbation can acquire a scale invariant spectrum from an initial matter-dominated phase of contraction.

\section{Non-Singular Matter Bounce Scenario}
\label{MatterBounce}

In this section, we elaborate upon how the S-brane mediated bouncing cosmology discussed in this paper can be used to provide an explicit non-singular realization of the ``matter bounce'' scenario. First, we briefly compare the essentials of the matter bounce and inflationary scenarios, after which we remind the reader how a matter/radiation system in a contracting universe can lead to a scale invariant spectrum of fluctuations which persists in the radiation/dilaton era, during which at some point the S-brane condenses. We then conclude that the late time power spectrum of curvature fluctuations is scale-invariant. 

\subsection{Matter bounce versus inflation}

The hypothesis of the ``matter bounce'' scenario is that the universe begins large and cold, after which matter-dominated contraction onsets during which all scales which are currently observed in cosmic microwave background (CMB) exit the Hubble radius. In Figure 1 we provide a spacetime sketch of the matter bounce scenario, which represents an alternative mechanism to inflation to generate cosmological perturbations \cite{Wands}. The vertical axis represents cosmological time $t$ and the horizontal axis, physical distances. The S-brane mediates a bounce at $t_c$, which can be chosen to be zero. Two length scales are indicated in Figure 1. First the physical wavelength $\lambda_{\rm ph}=2\pi a(t)/k$ of a fluctuation mode (solid line), second the Hubble radius $H^{-1}(t)$ (dashed line), which is linearly decreasing in the contracting phase and linearly increasing in the expanding period. It is important to stress here that since the energy density at the bounce point is finite, the minimal value of the Hubble radius is non-zero. This is a crucial property of our string cosmological scenario, which most other pre-Big Bang models do not possess. 
\begin{figure}[h!]
  \hfill
  \begin{minipage}[t]{.45\textwidth}
    \begin{center}
      \includegraphics[height=10cm]{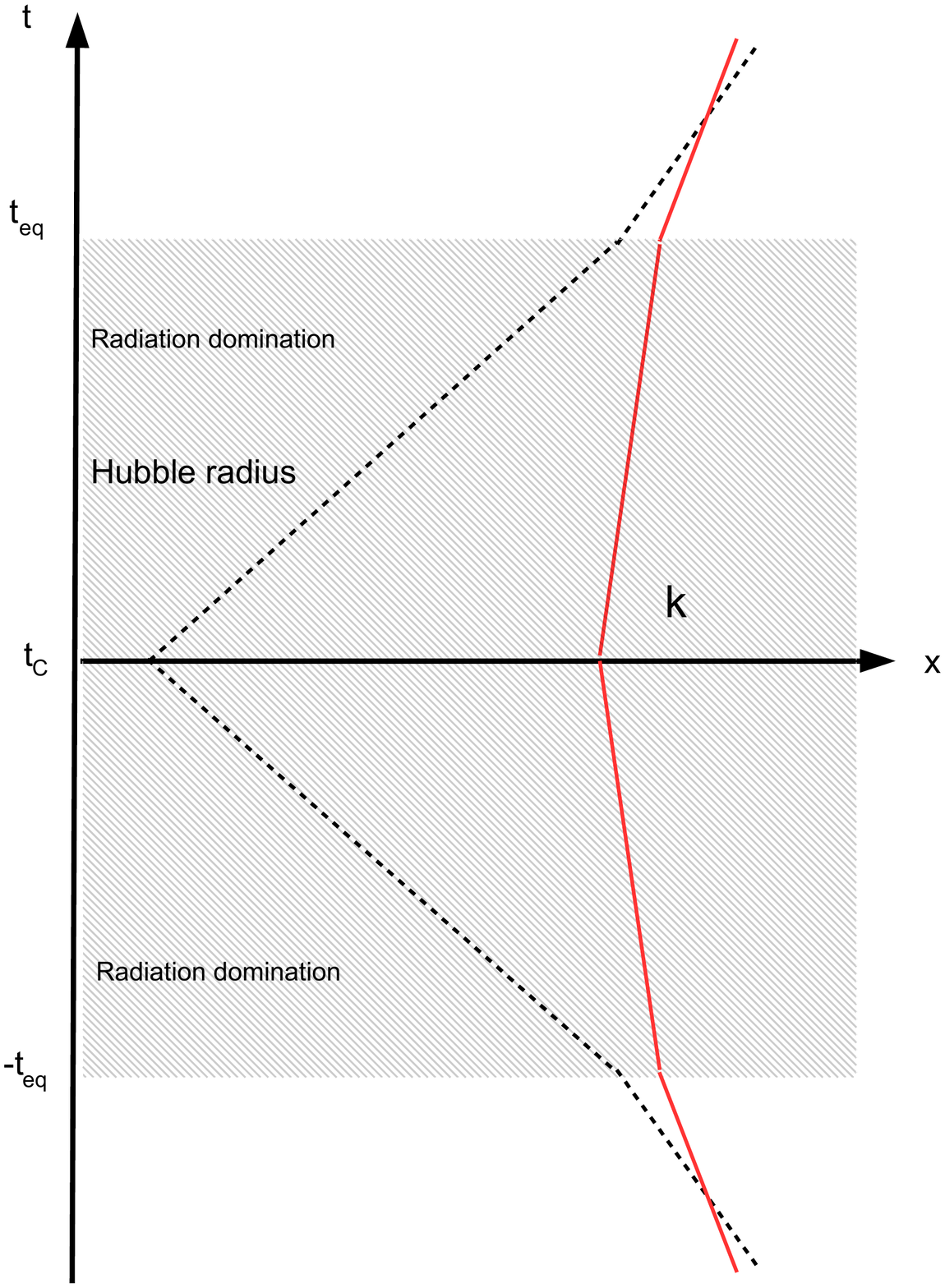}
      \caption{Spacetime diagram (sketch) of the matter bounce scenario. The vertical axis is cosmological time $t$, with $t_c=0$ indicating the bounce point. The horizontal axis represents physical length. The dashed line represents $H^{-1}$ and the bold  line is the physical wavelength of a fixed comoving scale $k$.}
      \label{cinin}
    \end{center}
  \end{minipage}
  \hfill
  \begin{minipage}[t]{.45\textwidth}
    \begin{center}
     \includegraphics[height=10cm]{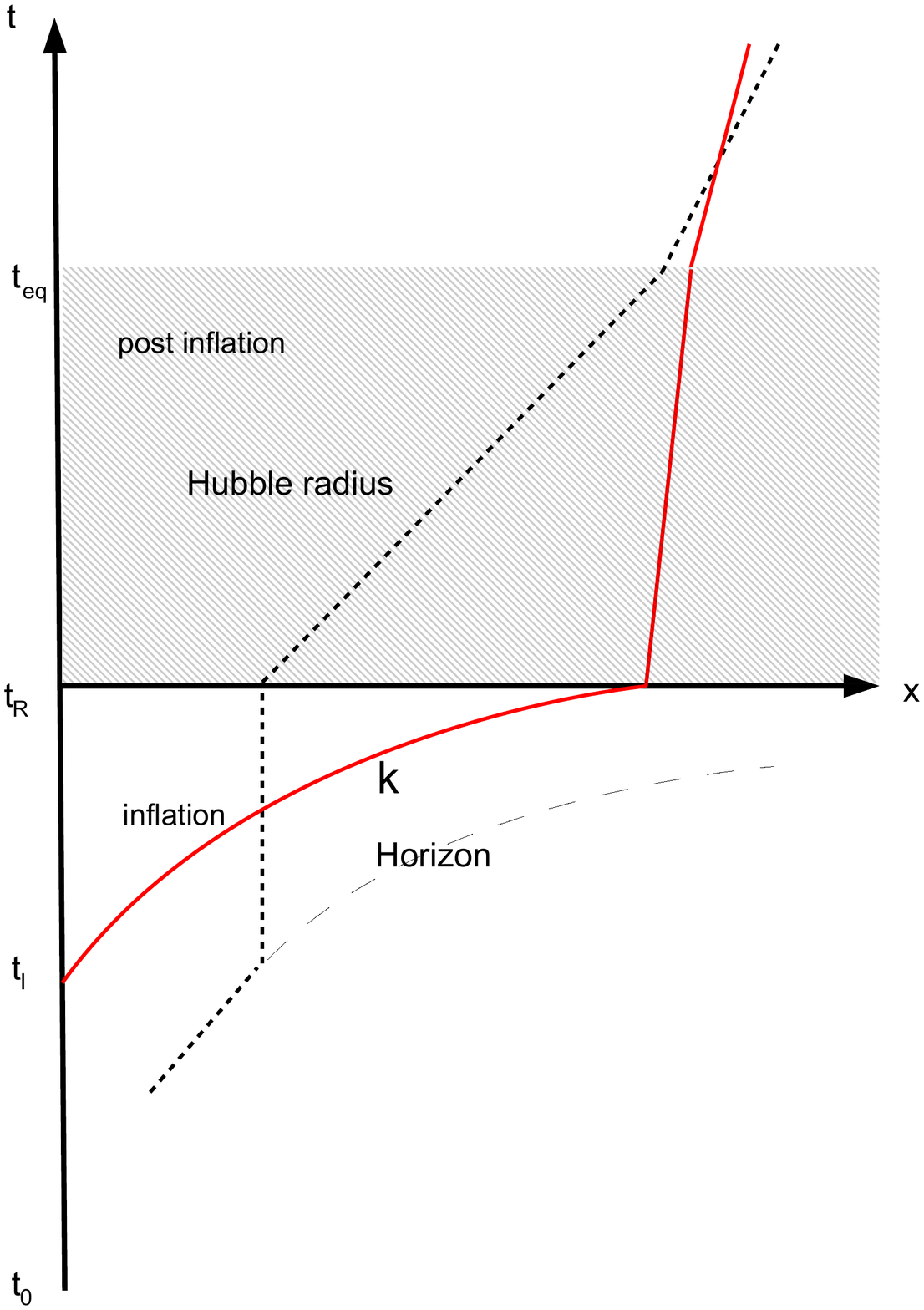} 
      \caption{Analogous diagram for inflationary cosmology, with $t_i$ denoting the beginning of the inflationary phase and $t_R$ the end. The particle horizon is the wide dashed curve.}
      \label{csm}
    \end{center}
  \end{minipage}
  \hfill
\end{figure}
We compare the spacetime diagram of the matter bounce to that of
inflationary cosmology\footnote{To be concrete, we assume almost
exponential expansion.}, sketched in Figure 2, on which we depict in addition the particle horizon (long dashed line) which expands exponentially during inflation.

Inflation is designed to solve the horizon, flatness and entropy problems of standard cosmology. The horizon problem is addressed by the fact that the horizon becomes exponentially larger than the Hubble radius during inflation. Conversely, the matter bounce scenario does not have a horizon problem since the horizon is always much larger than the relevant wavelengths\footnote{Note, in particular, that in the case 
that the contracting phase extends to $t = - \infty$, then the particle horizon is infinite. Even if the contracting period is finite, the horizon is much larger than the Hubble radius. }.

The exponential increase of the scale factor during the period of inflation dilutes spatial curvature and thus accounts for the observed flatness of
spatial sections, \textit{so long as the spatial curvature during the pre-inflationary phase is not too large to prevent the onset of inflation}. The matter bounce scenario mitigates the flatness problem. Unlike in any pre-Big Bang scenario with no period of inflation, the initial spatial curvature does not have to be tuned to an extremely small value to explain the current data. Initial flatness constraints at large radius in the contracting phase similar to the currently observed bounds on the spatial curvature are sufficient. Furthermore, there is no entropy problem in any pre-Big Bang scenario in which the universe starts out large.

The most striking success of inflationary cosmology concerns the prediction of an almost scale-invariant spectrum of curvature fluctuations. However, as already realized a decade before the development of inflationary cosmology \cite{Peebles, SZ}, sufficient conditions required to obtain curvature fluctuations in agreement with observations (in particular with the observed acoustic oscillations in the CMB angular power spectrum) is to have a primordial spectrum of coherent scale-invariant adiabatic fluctuations on super-Hubble scales at the time when the smallest observable scales enter the Hubble radius. To obtain this, it is necessary that the Hubble radius  in comoving coordinates ${\cal H}^{-1}$ decreases so that constant comoving scales $\lambda$ of interest emerge on sub-Hubble scales. 

Figures 1 and 2 show that this condition is satisfied both in
inflationary cosmology and in the matter bounce scenario. In the case
of inflation, the exponential increase of spatial sections causes physical fluctuation modes $\lambda_{\rm ph}$ to exit the Hubble radius $H^{-1}$. In the case of the matter bounce it is the contraction of space which leads to a decrease in the comoving Hubble radius. A second requirement is that the fluctuations are not sourced during the time interval in which the physical wavelength is larger than the Hubble radius. In this case, the wave function of fluctuation modes is squeezed as required in order to obtain acoustic oscillations in the CMB anisotropy spectrum. Thirdly, a mechanism is required to produce an almost scale-invariant spectrum. In the case of inflationary cosmology it is the time translation invariance of the inflationary phase which ensures this.
In the case of the matter bounce, as we will show in the following subsections, it is the specific growth rate of super-Hubble curvature fluctuations during the matter-dominated phase of contraction which transforms the initial vacuum spectrum into a scale-invariant one \cite{Wands, Fabio}.

\subsection{Matter bounce plus S-brane scenario}

The string cosmological model examined in this work must be completed in the low temperature regime, in both the expanding and contracting phases. As shown in previous work in the context of string models that exhibit spontaneous breaking of supersymmetry either via geometrical or non-geometrical fluxes,  during expansion the Universe gets attracted at lower temperatures to radiation-like, intermediate regimes
($\rho_{\rm total}=3p_{\rm total}$) with the supersymmetry breaking scale $M_{\rm susy}(t)$ evolving proportionally to the temperature $T(t)$  along a critical trajectory: $T(t)/M_{\rm susy}(t)$=constant \cite{stabmod,CriticalCosmo}.  
At sufficiently low temperatures, the supersymmetry breaking scale gets frozen and the relevant moduli including the dilaton get stabilized. The mass spectrum of the various moduli, including the no-scale modulus which controls the susy breaking scale, depends on the pattern of supersymmetry breaking \cite{Noscale}. There are moduli which acquire masses of the order of $\langle M_{\rm susy }\rangle \sim {\rm  {\cal O}(1)}$TeV, 
while other moduli typically get a mass of the order  of $\langle M^2_{\rm susy }\rangle/ M_{\rm Planck} \sim  {\rm  {\cal O}( 10^{-3})}$eV.
The light fields are those that participate in supersymmetry breaking, such as the no-scale modulus (the superpartners of the Goldstino combination). What is important for our purposes is that the dilaton acquires a mass as a result of susy breaking, and either possibility can be realized. For instance, in the breaking of susy via gaugino condensation \cite{gaugino} (for studies of moduli stabilization in the context of string gas cosmology see \cite{Xue,Patil:2004zp,Edna,Scott, Alex}) the dilaton gets a relatively high mass, $m_{\phi} \sim {\rm  {\cal O}(1)}$TeV, while in the cases of models where the supersymmetry breaking is based on stringy versions of the Scherk-Schwarz mechanism \cite{SS,SSstring}, the dilaton acquires a very light mass $m_{\phi} \sim{\rm  {\cal O}( 10^{-3})}$eV.

  
By requiring supersymmetry breaking and dilaton stabilization at lower energy scales, the string bouncing cosmology can be connected to early and late epochs of matter domination, thus providing a non-singular realization of the matter bounce scenario. This is because susy breaking generates particle masses such that at sufficiently low temperatures, the universe enters a matter dominated era with equation of state $p \simeq 0$.  The value of the dilaton mass is crucial as it leads to constraints on the parameters $c_{\sigma}$ and $c_\p$ controlling the fluctuations in the contracting thermal string gas/dilaton phase. We focus on cases where the dilaton acquires a mass of the order of  ${\rm  {\cal O}(1)}$TeV. We will comment on cases where the dilaton participates in the breaking of susy (and thus acquires a very small mass) at the end of this section, leaving a detailed
study of the resulting spectrum of cosmological fluctuations for this class of models for a future investigation. 

Bearing this in mind, we turn towards considering the evolution in the expanding phase. Once the temperature has dropped just below the dilaton mass, $m_{\phi} \sim {\rm  {\cal O}(1)}$TeV, the dilaton starts to roll towards its minimum, radiating its kinetic energy, eventually undergoing damped oscillations so that it rapidly freezes at its minimum. Thus the matter dominated era is preceeded by a radiation era where the dilaton is frozen. Note that the long wavelength fluctuations in this intermediate, radiation dominated phase are by default adiabatic since the dilaton is fixed and a single fluid dominates the evolution. 

By thermal duality symmetry, the cosmological regimes encountered in the expanding phase should appear also in the contacting phase but in reverse chronological order. Hence we begin with a large and asymptotically cold contracting phase where supersymmetry is broken, massive fields dominate the background and the dilaton is fixed. This is the initial matter-dominated phase of contraction. In this phase an almost pressureless hydrodynamical fluid dominates the background and sources fluctuations. We denote its energy density by $\rho_m$ and its pressure by $p_m=w\rho_m$, where $w$ is very small. There is also a subdominant radiation fluid component (with equation of state $\rho_r=3p_r$) which becomes relevant at around the time of equal matter and radiation denoted by $\tau_{\rm eq}$. The dilaton remains frozen at its minimum until the Universe has heated to the temperature corresponding to the supersymmetry breaking scale or the dilaton mass, whichever is earlier. The dilaton then becomes effectively massless and thus dynamical and the description of the fluids given in the earlier sections become applicable: The radiation dominated phase being admixed with the dynamical dilaton whose dynamics become more and more relevant as the S-brane bounce is approached. Taking the asymptotic value of the background dilaton field early on in the thermal string gas/dilaton phase to be given by the value of the dilaton at its minimum imposes that the coefficient $C_3$ vanishes, which in turn implies the relation
\be
C_3=0~~~\Longrightarrow ~~~c_\p=3c_\sigma ,  
\ee 
with $c_\p$ and $c_\sigma$ defined by Eqs (\ref{solsig}) and (\ref{solpp}). A consequence of this relation is that the evolution of the long wavelength fluctuations during the thermal string gas/dilaton phase including the transition to the expanding phase via the S-brane remains adiabatic throughout. As we will show below, the coefficient $c_{\sigma}$, which controls the asymptotic value of the gravitational potential $\Phi$, and hence the spectrum of curvature fluctuations will acquire a scale invariant spectrum from the initial matter phase of contraction. 

\subsubsection{Initial matter phase of contraction } 

We can now proceed to study the evolution of fluctuations in the initial matter dominated phase of contraction. As remarked earlier, in this phase the universe is dominated by a fluid with equation of state $p=w\rho$, with $w$ a small positive constant\footnote{Our analysis applies also in cases where the state equation coefficient $w$ is slowly varying with time, via the WKB approximation. For example, a ``baryon''-matter/radiation plasma, where the matter fluid and the radiation component do not decouple, gives rise to a situation where the coefficient $w$ is slowly varying and becomes very small in the matter dominated phase.}. The speed of sound, given by $c_s^2=w$ is correspondingly also small. As discussed in Appendix \ref{A}, a convenient way to describe the various states of the fluid matter is via a derivatively coupled scalar field $\psi_m$, in terms of which the pressure is written as follows:
\be
p=X^{1+w \over 2w}=\left(-{1\over 2}\partial_{\mu}\psi_m\partial^{\mu}\psi_m\right)^{1+w \over 2w}.
\ee
The background scale factor scales with conformal time $\tau$ according to the relation
\be
a\propto \abs\tau \abs^{2\over (1+3w)},   
\ee
while the background scalar potential $(\psi_m)_0$ is homogeneous having time dependence only. 

The fluctuations in the matter dominated phase are parametrized by the gravitational potential $\Phi$ and the fluctuation of $\psi_m$, which we denote by $\delta \psi_m$. In terms of these we introduce the gauge invariant, canonical variable 
\be
v=\sqrt{\rho_{,X}} a\left(\delta\psi_m+{(\psi_m^{\prime})_0\over \h}\Phi\right),
\ee 
whose modes satisfy the following equation of motion:
\be
v_k^{\prime\prime} +\omega^2(k) v_k=0\, , \quad \omega^2(k)=c_s^2k^2- {a^{\prime\prime}\over a}\, . 
\ee
Utilizing the expression for the background scale factor $a$, the frequency can be written as follows,
\be
\omega^2(k)=w k^2+\left[{1\over 4}-\nu^2\right]{1\over \tau^2}\, ~~~{\rm with}~~~\nu={3\over 2}\, {1-w\over 1+3w}  .
\ee
This equation admits exact solutions in terms of Bessel functions. The solution valid for all wavelengths is given by
\be
v_k=\sqrt{\abs\tau\abs}\left( B_+ J_{\nu} (\sqrt{w}k\abs\tau\abs)+B_- Y_{\nu}(\sqrt{w}k\abs\tau\abs)\right),~~~~\nu={3\over 2}\, {1-w\over 1+3w} \, ,
\ee
with $B_+$ and $B_-$ time-independent constants. 

At short wavelengths, where $\sqrt{w} k\abs \tau \abs \gg 1$, the asymptotic behavior of the solutions exhibits oscillatory behavior given by 
\be
 \sqrt{2\over \pi c_s k}\, \left[ B_{+} {\rm cos}\left (\sqrt{w}k\abs\tau\abs -{\pi\over2}\left(\nu+{1\over 2}\right)\right) + B_{-}\,{\rm sin}\left (\sqrt{w} k\abs\tau\abs -{\pi\over2}\left(\nu+{1\over 2}\right)\right)  \right]\, ,
\ee
showing that both modes are canonically normalized when $B_+$ and $B_-$ are $k$-independent constants. This oscillatory behavior at short wavelengths occurs independently of the equation of state coefficient $w$. It also arises in situations where the speed of sound is slowly varying with time, via the WKB approximation \cite{Mukhanovbook}. At short wavelengths the spectra of these two $v$-modes behave like that of vacuum fluctuations, which is blue.
  
At long wavelengths on the other hand, $\sqrt{w} k\abs \tau \abs \ll 1$, the asymptotic behavior of the Bessel functions leads to a power law behavior 
\be
v_k\sim \sqrt{\abs\tau\abs}\left[{B_+\over \Gamma(\nu+1)}\left({\sqrt{w}k\abs\tau\abs\over 2}\right)^{\nu}-B_ -{ \Gamma(\nu)\over \pi}\left({\sqrt{w}k\abs\tau\abs\over 2}\right)^{-\nu}\right].
\ee
Vacuum initial conditions for $v_k$  -- that is, $B_{\pm}$ are $k$-independent constants -- imply that the mode proportional to $B_-$ gives rise to an almost scale invariant spectrum when $\nu \simeq {3/ 2}-6w$ with $w$ small enough, while the $B_+$ mode gives rise to a blue spectrum which is irrelevant on super-Hubble scales.

Physically, this result can be understood as follows. The $B_-$ mode increases on super-Hubble scales. Since long wavelengths spend more time outside the Hubble radius, small $k$ modes are boosted relative to large $k$ modes, i.e. the spectrum reddens. What is particular about a matter-dominated phase of contraction is that the growing $B_-$ mode scales as $\tau^{-1}$, which is precisely the growth rate needed to convert the initially blue vacuum spectrum to a scale-invariant one. Thus, at the end of the matter-dominated phase of contraction we have
\be
\label{scav}
v_k \, \sim \,-B_-{ \Gamma\left({3\over 2}-6w\right)\over  \pi}~\abs\tau\abs^{-1+6w} \left({\sqrt{w}k\over 2}\right)^{-{3\over 2}+6w} \, ,
\ee
yielding a power spectrum
\be
P_v(k, \abs\tau\abs) \, = \frac{k^3}{12 \pi} |v_k|^2 \sim \, {2\over 3}\, {\Gamma\left({3\over 2}-6w\right)^2\over \pi^3 w^{3/2}}\, B_-^2\left({c_sk\abs\tau\abs\over 2} \right)^{12w}\,  \sim {\tilde B^2\over \tau^2}\left[1+12\, c^2_s\, \ln \left({c_sk\abs\tau\abs\over 2}\right) \right]\, ,
\ee
which is independent of $k$ up to small logarithmic corrections.

The almost scale invariant spectrum of the $v$ variable in the matter dominated phase of contraction will be inherited in the spectrum of curvature fluctuations in the ``baryon-radiation'' plasma epoch that follows just after. In order to see this inheritance realized, it will be more convenient to work in terms of the gravitational potential $\Phi$. First we work out the precise normalization of the $\Phi_k$ modes with respect to the co-moving wavenumber $k$, which follows from vacuum initial conditions in the matter dominated era. In this regime the $\Phi_k$ modes satisfy  the following equation:
\be
 \Phi_k^{\prime \prime}+{6(1+w)\over 1+3w}{1\over \tau}\Phi_k^{\prime} +wk^2\Phi_k=0,
 \ee
with the solution given in terms of Bessel functions
\be
  \Phi_k= {1\over \abs\tau\abs^{\alpha}}\left(C_+J_{\alpha}(\sqrt{w}k\abs\tau\abs)+C_-Y_{\alpha}(\sqrt{w}k\abs\tau\abs)\right),~~~\\ \alpha={1\over 2}\, {5+3w\over 1+3w}=\nu+1.
\ee
A very useful variable related to $\Phi$ is the Mukhanov variable $u$, 
\be
 u_k\equiv †{2\Phi_k\over \sqrt{\rho+p}}
\ee
which obeys a wave equation without friction term:
\be
 u_k^{\prime\prime} +\omega^2(k) u_k=0\, , \quad \omega^2(k)=w k^2+\left[{1\over 4}-\alpha^2\right]{1\over \tau^2}\, ,~~~\alpha={1\over 2}\, {5+3w\over 1+3w}\, .
\ee
This leads to the solution:
\be
  u_k=\sqrt{\abs\tau\abs}\left(\tilde C_+J_{\alpha}(\sqrt{w}k\abs\tau\abs)+\tilde C_-Y_{\alpha}(\sqrt{w}k\abs\tau\abs)\right),~~~\\ \alpha={1\over 2}\, {5+3w\over 1+3w}\, .
\ee
The time-independent constants $\tilde C_{\pm}$ are related to the $B_{\pm}$ constants \cite{Mukhanovbook}. Their precise relation follows from the  equation
\be
c_s\nabla^2 u=a\left({v\over a}\right)^{\prime}\, ,
\ee
which for short wavelenghts yields
$$
-c_sk^2\sqrt{2\over \pi c_s k}\, \left[\tilde C_{+} {\rm cos}\left (\sqrt{w}k\abs\tau\abs -{\pi\over2}\left(\alpha+{1\over 2}\right)\right) + \tilde C_{-}\,{\rm sin}\left (\sqrt{w} k\abs\tau\abs -{\pi\over2}\left(\alpha+{1\over 2}\right)\right)  \right]
$$
\be
=\sqrt{2\over \pi c_s k}\, \sqrt{w}k\left[B_{+} {\rm sin}\left (\sqrt{w}k\abs\tau\abs -{\pi\over2} \left(\nu+{1\over 2} \right) \right) -B_{-}\,{\rm cos}\left (\sqrt{w} k\abs\tau\abs -{\pi\over2}\left(\nu+{1\over 2}\right) \right)  \right]\, .
\ee
Using the fact that $\alpha=\nu+1 $, we obtain
\be
\tilde C_+=-{B_+\over  k}~,~~~~~~\tilde C_-=-{B_-\over   k}\, .
\ee
In contrast to the $B_{\pm}$ modes,  the $\tilde C_{\pm}$ and $C_{\pm}$ modes acquire a $k$-dependence via a factor of $k^{-1}$. This fundamental relation between the modes of $v$ and the modes of $u$ (or $\Phi$) is universal, holding for any equation of state, as well as when the speed of sound varies slowly enough (defined by the smallness of the parameter $s = \dot c_s/(H c_s)$).  

In the long-wavelength limit, the properly normalized $u_k$ modes behave as:
\be
u_k\sim \sqrt{\abs\tau\abs}\left[{\tilde C_+\over \Gamma(\alpha+1)}\left({\sqrt{w}k\abs\tau\abs\over 2}\right)^{\alpha}-\tilde C_ -{ \Gamma(\alpha)\over \pi}\left({\sqrt{w}k\abs\tau\abs\over 2}\right)^{-\alpha}\right] ~~{\rm with} ~~\alpha\simeq {5\over 2}-6w\, .
\ee
Therefore, the $\tilde C_-$  growing mode scales as $k^{-7/2}$ with the co-moving wavenumber. This precise scaling, following from the initial conditions in the matter dominated phase fixes the $k$-dependence of the $u$ (and also the $\Phi$) modes for their subsequent evolutions. As we demonstrate in the following subsection, the $\tilde C_-$  mode induces a scale invariant spectrum for the coefficient $c_{\sigma}$ in the radiation phase which controls the constant mode of the curvature fluctuation throughout the cosmological evolution, even after propagating across the brane into the expanding phase.

\subsubsection{Matter and radiation plasma}

We proceed now to review the evolution of cosmological fluctuations during the contracting matter/radiation era following \cite{MFB, Mukhanovbook}. The background scale factor and energy densities are given by
\be
a(\tau)=a_{\rm eq}\chi(\chi+2)\, , ~~~{\rm with}~~~\chi\equiv {\tau \over \tau_*},~~~\tau_*={\tau_{\rm eq}\over \sqrt{2}-1}
\ee
\be
\rho_m={\rho_{\rm eq}\over 2}\left(a_{\rm eq}\over a\right)^3,~~~\rho_r={\rho_{\rm eq}\over 2}\left(a_{\rm eq}\over a\right)^4,
~~~\tau_*=-\sqrt{24\over \rho_{\rm eq}a_{\rm eq}^2 } \, ,
\ee
where $a_{\rm eq}$ and $\rho_{\rm eq}$ are the scale factor and total energy density at the time of equal matter and radiation $\tau_{\rm eq}$ (which in the contracting phase we take to be negative). The speed of sound is given by
\be
c_s^2={1\over 3}\left(4\rho_r \over  3\rho_m +4\rho_r \right)={1\over 3}\left({4\over 3\chi(\chi+2)+4}\right).
\ee
At early negative  times,  $\abs \tau\abs \gg \abs \tau_{\rm eq}\abs$, when the matter fluid dominates, the speed of sound tends to zero, while at times
$\abs\tau\abs \ll \abs\tau_{\rm eq}\abs$, it acquires the constant value of 
$1/3$.

In order to relate the parameters of the initial matter dominated regime with the parameters of the matter/radiation plasma era, it is convenient to work with the gauge invariant quantity $\zeta$. In terms of the variable $u$, this quantity is given by
\be 
\zeta={\rm  sign}(\h)~ \theta^2\left( u\over\theta \right)^{\prime},~~{\rm where}~~ \theta={1\over \sqrt{3} \, a} \left({\rho \over \rho +p}\right)^{1\over2} ={(\chi+1)\over   a_{\rm eq}\,\chi(\chi+2)\sqrt{3\chi(\chi+2) +4}}
\,.
\ee
The above expression for $\zeta$ is valid both at short and long wavelengths. 
We are interested to examine the transition of $\zeta$ from the initial matter dominated regime to the radiation era, 
and understand how the relevant modes behave at long wavelengths. To this end, it is sufficient to consider the long wavelength solution, keeping however the leading $k^2$-corrections via the integral equation for $u$ \cite{Mukhanovbook}:
\be
u_k=u_k^0-k^2\theta \int^\tau{d\tau^{\prime} \over \theta^2}\left(\int^{\tau^\prime} d\tau^{\prime \prime}c^2_s\theta u^0_k\right), ~~{\rm where }~~u^0_k=\hat C_-\theta+\hat C_+\theta \int^\tau {d\tau^{\prime}\over \theta^2} \, .
\ee
The $k^2$-corrections turn out to be key for our considerations. Indeed neglecting them, one would obtain
\be
\zeta^0=-\, \hat C_+ \, ,
\ee 
without any dependence on the relevant $\hat C_-$ mode of the matter dominated regime. Even worse, the $\hat C_+$ mode is irrelevant on super Hubble scales as it gives rise to an ultra blue spectrum. Therefore, the relevant contributions to $\zeta$ arise from the $k^2$-terms, and more specifically from terms proportional to the $\hat C_-$ mode. Neglecting the irrelevant blue modes, one finds that in the radiation dominated regime, $\zeta$ is given by:
\be
\zeta(\tau_r)= k^2\hat C_-\int_{\tau_m}^{\tau_r}d\tau c_s^2\theta^2\, = - k^2\hat C_-\, \abs\tau_*\abs \int_{\chi_m}^{\chi_r}d \chi \, c_s^2\theta^2\, ,
\ee
where the limits of integration must satisfy the following inequalities: $\abs\tau_m\abs\gg \abs\tau_{\rm eq}\abs$ and $\abs \tau_r\abs \le\abs\tau_{\rm eq}\abs$, and so $\chi_m\gg 1$  and $\chi_r \simeq 1$.   The lower integration limit ensures overlap of the solution with that of the initial matter dominated regime (analyzed in the previous subsection), implying that the scaling of the coefficient $\hat C_-$ with $k$ is the one determined by the vacuum initial conditions. The smallness of $c_s$ in the matter dominated regime imposes that $\abs\tau_m\abs\gg \abs \tau_{\rm eq}\abs$. The upper limit is conveniently chosen to lie in the radiation dominated regime, and provides the initial condition for the curvature perturbation for the subsequent cosmological evolution.  
 
As already demonstrated in Section \ref{ZETA}, upon taking into account that the coefficient $C_3$ must vanish, from the onset of the radiation regime and during the subsequent cosmological evolution, the curvature perturbation $\zeta$ remains constant and is equal to the fundamental parameter $-c_{\sigma}$. This parameter controls the evolution of the fluctuations at latter cosmological times, even after the S-brane crossing to the expanding phase. The coefficient $C_3$ is zero since the dilaton is frozen to its minimum during the radiation dominated era, before the thermal string gas/dilaton fluid dominance. These considerations impose that $-c_\sigma = \zeta(\tau_r)$. The scaling of the coefficient $\hat C_-$ with the wavenumber $k$ was determined in the initial matter dominated regime and turns out to scale like $k^{-7/2}$. This is the precise scaling needed to render the power spectrum of $\zeta_k(\tau_r)$, and hence the spectrum of $c_\sigma$, scale invariant. Indeed, the curvature perturbation at the onset of the radiation era is given by
\be
-c_\sigma(k)=\zeta_k(\tau_r) = C_\zeta \, k^{-{3 \over 2}}\, ,
\ee
where the coefficient $C_\zeta$ is $k$-independent, showing that the power spectrum of the curvature perturbation $c_\sigma$ is scale invariant:
\be
P_{c_\sigma} \, = \frac{k^3}{12 \pi} |\zeta_k(\tau_r)|^2 = {1\over 12\pi}\abs C_\zeta \abs^2 \, .
\ee
Therefore the scale invariant spectrum of the growing mode in the initial matter dominated contracting phase is inherited by the constant mode of the curvature fluctuation through the subsequent cosmological evolution and is preserved after the S-brane mediated bounce on to the expanding phase. The above scaling property can also be found from the initial matter dominated phase by noticing that in this regime, 
\be
\zeta_k = -c_s\theta v_k\, ,
\ee
which shows that the coefficient $k^2\hat C_-$ behaves as $B_- k^{-3/2}$  (see Eq. (\ref{scav})).  

We conclude this section with some remarks concerning the other class of string models exhibiting spontaneous breaking of susy, namely models where the dilaton participates in the susy breaking mechanism, and thus acquires a very small mass: $m_{\phi}=\langle M^2_{\rm susy }\rangle/ M_{\rm Planck} \sim  {\rm  {\cal O}( 10^{-3})}$eV. In this case the dilaton becomes effectively massless during the initial matter dominated phase of contraction, and thus dynamical thereafter,  giving rise to a non-zero coefficient $C_3$. As can be seen from Eq. (\ref{zeta}), for this case the long wavelength perturbation $\zeta$ is not conserved during the thermal string gas/dilaton phase, implying the presence of non-adiabatic, isocurvature modes. The contribution of these modes acquires its maximal value at the S-brane, but it is negligible at lower temperatures, as it is suppressed by a factor of $(\tilde T / \tilde T_c)^2$. There is however a very interesting, non-trivial effect in that the dilaton admixture to the matter fluid causes the effective speed of sound to change, giving rise to a tilt in the spectral index of curvature fluctuations. Qualitatively, this tilt, which depends on the dilaton mass $m_{\phi}$, will appear as a violation of exact scale invariance for the spectrum. The existence of such a tilt is supported by recent observational data. The complete analysis of this case is rather involved, as it requires to investigate the cosmological fluctuations in a three fluid system, however it would be very interesting to carry it out in order to compare the predictions of the model for spectral index with the observational data. We plan to quantitatively analyze such string models in future work.

\section{Conclusions}

We have studied the transfer of cosmological fluctuations through a stringy S-brane, which mediates the transition between contracting and expanding cosmological phases related via string thermal duality symmetry. The S-brane itself is a space-filling defect that interpolates between the two dual geometrical phases of the underlying worldsheet CFT, and is sourced by non-trivial thermal string states that become massless at a critical maximal temperature. As we demonstrated in this paper, the energy density of the spacelike brane vanishes, {\it even in the presence of metric and dilaton fluctuations}, while the pressure along the spatial dimensions is negative. Therefore, the brane provides the violation of the NEC needed to induce a transition from contraction to expansion (in the Einstein frame). Since the S-brane has a string scale thickness in time, it is realized instantaneously from the perspective of the low energy effective theory. 

The resulting spacelike hypersurface is distinct, and has not been considered previously in the literature. It is defined as a surface of constant string frame temperature, with the latter reaching its maximal value as imposed by the thermal duality symmetry of the underlying string theoretic model. From the Einstein frame point of view, the S-brane surface is not isothermal since the physical temperature is dressed by a non-trivial dilaton factor. It does not correspond to a surface of constant energy density either. These properties turn out to be crucial for the matching of cosmological fluctuations across the bounce. Two of our key results is that there is no thermal entropy production during the stringy transition, and that the dilaton field, including its fluctuation, is mirror reflective across the brane.

Working in longitudinal gauge, we followed the evolution of the gravitational  potential $\Phi$ and that of the dilaton fluctuation $\p$. A crucial property of the underlying string cosmological background solutions is that the Einstein frame Hubble parameter $\h$ does not vanish at the bounce (where the S-brane materializes). This is because the dilaton field has a non-trivial kinetic energy density which is conserved as the dilaton bounces elastically across the brane. This is another distinguishing feature of the string cosmological solutions under consideration compared to other models in the literature. In each cosmological phase there are two independent modes for the gravitational potential $\Phi$ and for the dilaton fluctuation $\p$. In the expanding phase, the dominant modes of both $\Phi$ and $\p$ at super-Hubble scales are constant, while the sub-dominant modes are decaying in time. In the contracting phase there are in addition to constant long wavelength modes, growing modes that become relevant as the S-brane is approached. Due to the underlying thermal duality symmetry, the constraint of a maximal string frame temperature and the conservation of thermal entropy, the growing modes never become dominant. Their amplitudes are bounded never exceeding those of the constant modes. This is in contrast to the situation arising in other pre-Big Bang proposals. 
     
One of our main results is that both the constant modes and growing modes participate in the junction conditions across the S-brane hypersurface. The fact that the string frame temperature attains its maximal value at the bounce is crucial in implementing the junction conditions consistently. We show that the junction conditions are compatible with the Israel matching conditions in the Einstein frame. A very important consequence of our analysis is that there is a non-trivial coupling between growing modes in the contracting phase and the constant mode in the expanding phase that controls the strength of the late times curvature fluctuation. This contrasts to what is obtained in toy models of non-singular bounces \cite{quintom, HLbounce} where one finds highly suppressed couplings. The important lesson we draw from our work is thus that the details of how the transition between the contracting and expanding phases is achieved is key to understanding how the cosmological fluctuations propagate from one phase to the other. This sensitivity to the details of the mechanism which produces the bounce is not unexpected since it was already identified in different contexts in \cite{MP, DV}.

Having in our disposal a non singular, bouncing string cosmology as well as having correctly matched the cosmological fluctuations across the S-brane, we proceeded to consider the realization of the {\it matter bounce} scenario within our stringy framework. To complete the cosmological evolution at very low temperatures in both the contracting and expanding phases, we needed to implement a supersymmetry breaking mechanism. To this end, we distinguish between two generic classes of models. In the first class the dilaton does not participate in the susy breaking mechanism, receiving a relatively high mass of the order of the susy breaking scale $M_{\rm susy}$. In the second class, the dilaton is assumed to participate in the susy breaking mechanism thus acquiring a characteristic mass of order $M_{\rm susy}^2/M_{\rm Planck}$. As a result prior to the contracting thermal string gas/dilaton phase, there is an earlier phase in which supersymmetry is broken. In the cases where the dilaton acquires a relatively high mass, the dilaton is frozen and masses are generated, yielding an initial matter-dominated contracting phase with a subdominant radiation component that becomes relevant at a time of equal matter and radiation. A consequence of dilaton stabilization with a mass of the order of $M_{\rm susy}$ (and the matching conditions at the S-brane) is that the evolution of the long wavelength perturbations during the subsequent thermal string gas/dilaton phase and on to the expanding phase through the bounce is adiabatic. 

To avoid gauge ambiguities we work in terms of the Mukhanov-Sasaki canonical variable $v$. This variable inherits a scale-invariant power spectrum in the initial matter dominated phase of contraction. We showed that this scale invariant spectrum is  communicated to both modes of the gravitational potential $\Phi$ in the contracting phase, at the onset of radiation dominance. On scales of cosmological interest today, which have exited the Hubble radius long before the supersymmetry restoring phase transition, the slope of the spectrum of both modes is maintained, since all modes are evolving for the same amount of time in this phase and the change in the amplitude only depends on the time interval spent in this phase\footnote{Recall that the reason why a vacuum spectrum is converted to a scale-invariant one during the matter phase of contraction is that long wavelength modes spend more time in this phase on super-Hubble scales than short wavelength ones.}. The matching conditions at the stringy S-brane then show that the scale-invariance of the dominant mode of curvature fluctuations at late times in the expanding phase is inherited from the spectrum of the two $\Phi$-modes before the bounce.

Therefore the string bouncing cosmology considered in this work provides a successful, non-singular realization of the matter bounce scenario. Another advantage of our scenario compared to other bouncing models studied in the literature is that the bounce is not mediated by ad hoc new physics constructions such as ghost condensates or higher derivative gravity actions. Furthermore, in the model studied in this investigation there is no danger of new unstable modes appearing in the bounce phase. This problem, stressed recently in \cite{Paul} in the context of the ``New Ekpyrotic scenario" \cite{NewEkp} cannot arise in our setup\footnote{It is worth while pointing out that this problem is not generic to all effective field theory bouncing models. For example, the model of \cite{CEB} does not suffer from this problem.}.

We take note of the potential issue of the instability of a contracting phase to the growth of anisotropies -- the famous BKL instability \cite{BKL}. This is a problem faced in all bouncing models known to us with the exception of the Ekpyrotic scenario \cite{Ekp} and models \cite{CEB} using an Ekpyrotic phase of contraction to supplement the matter bounce scenario. The reason the latter models evade the BKL instability is that the field responsible for the contracting phase has an energy density which blueshifts faster than the anisotropic contributions to the energy density (which blueshift as $\sim 1/a^6$), and so negates the influence of the latter. In our model, we find ourselves in a marginal situation, wherein the dilaton kinetic energy blueshifts identically to the contributions to the energy density from anisotropy. Thus in addition, we have to assume that we have initial conditions such that the initial energy density of the dilaton as we begin the radiation dominated contraction dominates that of any initial anisotropies.

Even though we did not carry out a quantitative analysis for the cases with a very light dilaton mass, we argued that the dilaton motion during the initial matter dominated phase of contraction will produce a small violation to exact scale-invariance for the spectrum of curvature fluctuations, or a tilt to the spectral index. A second effect is the generation of initial ``entropy'' modes. One could worry that these modes could dominate the spectrum of the final curvature fluctuations (in the context of inflationary cosmology this is the {\it curvaton} mechanism \cite{curvaton}). However as we have seen in our work, the analysis of such modes in the long wavelength approximation during the thermal string gas/dilaton expanding phase reveals that at latter times their contribution to the curvature fluctuation is highly suppressed by a factor of $(\tilde T / \tilde T_c)^2$. 

The issue of obtaining a realistic spectrum that is not scale invariant is an interesting one. Given the current CMB data from the South Pole Telescope \cite{SPT}, the Atacama Cosmology Telescope \cite{ACT} and the Planck mission \cite{Planck}, a small red tilt is strongly favored. Therefore it is imperative to carry out the complete analysis of the light dilaton cases in order to investigate whether its predictions can explain the observational data. However we believe the correct perspective to have of the positive result of our investigation is that of a proof of concept. With this, one could be emboldened to look for concrete constructions that can more realistically model the late time universe in which we find ourselves.

\begin{acknowledgments} 

We are grateful to L. Alvarez-Gaum\'e, C. Bachas, R. Durrer, I. Florakis, A. Gaumez, J.-O. Gong, D. Luest,  P. Steinhardt, J. Troost, G. Veneziano, F. Vernizzi,  and especially S. Mukhanov  for fruitful discussions. RB, CK and HP wish to thank the CERN Theory Division for hospitality and support during the period when this project was begun. RB, HP, SP and NT wish to thank the Laboratoire de Physique Th\'eorique of the Ecole Normale Sup\'erieure for hospitality during visits while the project was under way. CK and HP would like to thank the University of Cyprus for hospitality. SP and NT would like 
to thank the Centre de Physique Th\'eorique of Ecole Polytechnique for hospitality. The work of CK and HP is partially supported by the ANR 05-BLAN-NT09-573739,  IRSES-UNIFY. The work of 
CK, HP, SP and NT was or is supported by the CEFIPRA/IFCPAR 4104-2 project and a PICS France/Cyprus. The work of HP is partially supported by the  EU contracts PITN GA-2009-237920, ERC-AG-226371 and PICS France/Greece, France/USA. The research of RB has been supported by an NSERC Discovery Grant, by the Canada Research Chairs Program and by a Killam Research Fellowship.  SP is supported at CERN by a Marie Curie Intra-European Fellowship of the European Community's 7'th Framework Program under contract number PIEF-GA-2011-302817.
\end{acknowledgments} 

\appendix

\section{Perturbing the string thermal fluid -- the thermal potential approach}
\label{A}
For a barotropic fluid (i.e. one for which $\rho = \rho(p)$ alone), 
one scalar potential suffices to determine the various states of the fluid. 
As demonstrated in \cite{Mukhanovbook,Dubovsky:2005xd} and reviewed in \cite{fv}, 
any perfect fluid with the equation of state $\rho = \gamma p$, with constant $\gamma$,  can be modeled by the action of a derivatively coupled scalar 
field $\psi$ with action
\eq{dca}{S = \int d^4x\sqrt{-g} \Bigl(-\frac{1}{2}\partial_\mu \psi \partial^\mu\psi \Bigr)^{\gamma +1\over 2} \, .}
To see this, we first consider the more general expression
\eq{dca2}{S = \int d^4x\sqrt{-g} P(X), ~~~{\rm with}~~~X= -\frac{1}{2}\partial_\mu\psi\partial^\mu\psi \, ,}
the form of which has been invoked to model D-brane moduli dynamics \cite{dbi}, kinetic models of inflation \cite{kinfl} and quintessence \cite{kess} as well as single field effective theories up to quartic order in derivatives \cite{cliff}. The equation of motion that results is
\eq{xdefeom}{\frac{1}{\sqrt{-g}}\partial_\mu(\sqrt{-g}P'g^{\mu\nu}\partial_\nu\psi)=0,}
and the energy momentum tensor is given by
\eq{kemt}{T^{\mu\nu} = g^{\mu\nu}P + P'\partial^\mu\psi\partial^\nu\psi \, .}
Comparison to the energy momentum tensor of an irrotational fluid
\eq{pfemt}{T^{\mu\nu} = g^{\mu\nu}p + u^\mu u^\nu(\rho + p)}
implies 
\be
\label{pdefk}
p=P\, ,~~~~\rho + p = 2XP' \, ,~~~~~~~u^\mu =\frac{\partial^\mu\psi}{\sqrt{-\partial_\nu\psi\partial^\nu\psi}},
\ee
whence one straightforwardly sees that if $P = X^{\gamma+1\over 2}$, then (\ref{pdefk}) 
implies that a fluid with the equation of state $\rho= \gamma p$ is to be described by a derivatively 
coupled scalar field $\psi$, as in (\ref{dca}). Furthermore, the equation 
of motion (\ref{xdefeom}) implies that
\eq{consvec}{J^\mu = P'\sqrt{2X}u^\mu}
is a conserved current. Comparing with the continuity equation of a compressible fluid
\eq{cecf}{\nabla_\mu(n u^\mu)=0\, ,}
where $n$ is the particle number density, we identify
\be
n = P'\sqrt{2X}={\rho + p \over \sqrt{2X}}\, .
\ee
Given that the thermal action 
\eq{sfact2}{S_T = \int d^4x\sqrt{-g}\, \Lambda T^4}
describes a fluid with equation of state $\rho=3p$, we conclude that $\sqrt{2 X}$ is to be identified with the temperature. 
In other words we
may introduce the {\it thermal potential} $\psi$ in terms of which the temperature is given by
\be
T = \sqrt{-\partial_\mu\psi\partial^\mu\psi}\, ,
\ee
and the action 
\eq{dcaf}{S_T =\int d^4x \sqrt{-g}\, \Lambda(-\partial_\mu\psi\partial^\mu\psi)^2 .}
Evidently the density $n$ corresponds to the thermal entropy density
\be
s={\rho + p \over T}\, .
\ee

\subsection{The temperature fluctuations away from the S-brane}
Perturbing the thermal action to second order, with $h_{\mu\nu}$ the metric perturbation and 
with $\psi = \psi_0 + \sigma$, results in
$$
S_T^{(2)} = \Lambda \int dx^4 \sqrt{-g}\, \left\{\Bigl(\frac{h^2}{8} - \frac{h_{\lambda\beta}h^{\lambda\beta}}{4}\Bigr)(\nabla_\mu\psi_0\nabla^\mu\psi_0)^2 - hh^{\mu\nu}\nabla_\mu\psi_0\nabla_\nu\psi_0\nabla_\kappa\psi_0\nabla^\kappa\psi_0 \right\}
$$
$$
 + 2\Lambda \int dx^4 \sqrt{-g}\,\Big\{h\nabla_\kappa\psi_0\nabla^\kappa\psi_0\nabla_\mu\psi_0\nabla^\mu\sigma +h^{\mu\lambda}h_{\lambda}^\nu\nabla_\mu\psi_0\nabla_\nu\psi_0\nabla_\kappa\psi_0\nabla^\kappa\psi_0 \Big\}
 $$
 $$
 +\Lambda \int dx^4 \sqrt{-g}\, \Big\{ h^{\mu\lambda}h^{\kappa\beta}\nabla_\mu\psi_0\nabla_\lambda\psi_0\nabla_\kappa\psi_0\nabla_\beta\psi_0 - 4 h^{\mu\nu}\nabla_\mu\sigma\nabla_\nu\psi_0\nabla_\kappa\psi_0\nabla^\kappa\psi_0 \Big\}
 $$
 \be
\label{aso2}
+\Lambda \int dx^4 \sqrt{-g}\, \Big\{   - 4 h^{\mu\nu}\nabla_\mu\psi_0\nabla_\nu\psi_0\nabla_\kappa\psi_0\nabla^\kappa\sigma + 2\nabla_\mu\sigma\nabla^\mu\sigma \nabla_\kappa\psi_0\nabla^\kappa\psi_0
  +4\nabla_\mu\sigma\nabla^\mu\psi_0 \nabla_\kappa\sigma\nabla^\kappa\psi_0\Big\} \, .
\ee
The perturbed energy momentum tensor is obtained as
\begin{eqnarray}
\label{pemtsf}
\delta T^\mu_\nu = \delta \Bigl[\frac{2}{\sqrt{-g}}g_{\nu\lambda}\frac{\delta S_T}{\delta g_{\lambda \mu}}\Bigr]&=&  -\frac{h}{2}T^\mu_\nu + h_{\nu\lambda}T^{\lambda\mu} +  \frac{2}{\sqrt{-g}}g_{\nu\lambda}\frac{\delta^2 S_T}{\delta g_{\lambda \mu} \delta g_{\kappa\tau}}h_{\kappa\tau} \nonumber + \frac{2}{\sqrt{-g}}g_{\nu\lambda}\frac{\delta^2 S_T}{\delta g_{\lambda \mu}\delta \psi}\delta\psi
\end{eqnarray}
from which one can read off the components
\eq{p0i}{\delta T^0_i = -\frac{4\Lambda\psi_0'^4}{a^4~}~\frac{\partial_i\sigma}{\psi_0'}}
\eq{pii}{\delta T^i_i = \frac{4\Lambda \psi_0'^4}{a^4~}~\Bigl(\frac{\sigma'}{\psi_0'} - \Phi\Bigr)}
\eq{p00}{\delta T^0_0 = \frac{-12\Lambda \psi_0'^4}{a^4}~\Bigl(\frac{\sigma'}{\psi_0'} - \Phi\Bigr) \, ,}
with $\Phi$ defined as in (\ref{lp}). Furthermore, varying (\ref{aso2}) to second order in the 
velocity potential results in the relativistic diffusion of temperature inhomogeneities via 
\eq{vpp}{\sigma'' -\frac{1}{3}\nabla^2\sigma + 2\frac{\psi_0''}{\psi_0'}\sigma' - 2\Phi'\psi_0' - 6\Phi\psi_0''= 0\, .}
It helps to recast the above in more familiar terms by accounting for the definition of temperature in terms of $\psi$, 
which perturbed to first order in the gauge defined by (\ref{lp}) implies
\eq{tpdef}{\frac{\delta T}{T_0} = \frac{h_{00}}{2a^2} + \frac{\sigma'}{\psi_0'} = -\Phi + \frac{\sigma'}{\psi_0'}.}
Given (\ref{pdefk}) applied to our example with $P = \Lambda X^2$ in conjunction 
with (\ref{tpdef}) implies 
\eq{p0if}{\delta T^0_i = -(\rho+p)\frac{\partial_i\sigma}{\psi_0'}}
\eq{piif}{\delta T^i_i = 4p \frac{\delta T}{T_0} = -4p \frac{\delta \beta}{\beta_0}}
\eq{p00f}{\delta T^0_0 = -12p \frac{\delta T}{T_0} = 12p \frac{\delta \beta}{\beta_0} \, .}
Although the latter two perturbed components of the energy momentum tensor may seem 
to obviously result from perturbing (\ref{bga}) with respect to $T$, (\ref{p0if}) is not so obvious. 
In order to perturb a system it is important to identify the fundamental degrees of freedom that 
are to be perturbed, and not the macro-states which describe equilibrium configurations of the 
system. The thermal potential description correctly accounts for the propagating 
degrees of freedom for a barotropic fluid. 

\subsection{The S-brane action in terms of the thermal potential}
\label{A2}

The scope of this section is to derive the exact functional form of the S-brane action in terms of the full metric and the thermal potential $\psi$. 
The generic S-brane action is written in terms of the induced metric $\gamma_{ab}$:
\be
\gamma_{ab}=g_{\mu\nu} \frac{\partial X^\mu}{\partial \xi^{a} }\frac{\partial X^\nu}{\partial \xi^{b}}\, ,
\ee
where the coordinates of the brane are denoted by $\xi^{a}$ and $X^{\mu}$ are the spacetime embedding fields: $X^{\mu}=X^{\mu}({\xi^{a}})$. 
We can always choose a gauge where the spatial coordinates are identified with those of the S-brane:
\be
X^{i}: =x^i=\delta^i_{a}\xi^{a}\,  .
\ee
Assuming that $g_{\mu\nu}$ is block-diagonal, $g_{i0}=g_{0i}=0$, the induced metric is given by
\be
\gamma_{i j}=g_{ij} +g_{00}~\partial_i X^0 \partial _j  X^0\, .
\ee
It is then clear that the function $X^0(x^i)$ specifies the S-brane embedding into spacetime.

Our goal is to specify the functional form of the embedding which is relevant to the isothermal surface. Notice that the determinant of $\gamma_{ij}$ is given by:
\be
{\rm det}(\gamma_{ij})={\rm det}(g_{ij})\left(1+g_{00}~ g^{ij}\partial_i X^0 \partial _j  X^0  \right )={\rm det}(g_{\mu\nu})\left(g^{00}+ g^{ij}\partial_i X^0 \partial _j  X^0  \right )
\ee
or
\be
{\rm det}(\gamma_{ij})=-{\rm det}(g_{\mu\nu})\left(-g^{00}\partial_0X^0 \partial _0  X^0 -g^{ij}\partial_i X^0 \partial _j  X^0  \right )
=-{\rm det}(g_{\mu\nu})\left(-g^{\mu \nu}\partial_{\mu}X^0 \partial _{\nu}  X^0  \right ).
\ee
The expression above is written in the special frame where the time-direction is orthogonal to the S-brane.

In order to make the derivation of the brane action $S_B$  more transparent, we first examine the case of a conformally flat metric. In this case we may always decompose the metric in terms of the time-like velocity $u_{\mu}=(1,0,0,0)$: 
$$
ds^2=e^{2\omega}\eta_{\mu\nu}dx^{\mu}dx^{\nu} = e^{2\omega}\left[-u_{\mu} u_{\nu} dx^{\mu}dx^{\nu} +
(\eta_{\mu \nu} +u_{\mu} u_{\nu} )dx^{\mu}dx^{\nu} \right] = e^{2\omega}\left[-u_{\mu} u_{\nu} dx^{\mu}dx^{\nu} +
\eta_{ij }dx^{i}dx^{j} \right] \, .
$$ 
In more general cases, 
the decomposition of the metric in terms of a generic velocity vector and in terms of a three dimensional subspace orthogonal to $u^{\mu}$ admits the following form:
\be
ds^2=-N^2\psi_{\mu} \psi_{\nu} dx^ {\mu}dx^{\nu}  +g_{ij}dx^{i}dx^{j}, ~~~~~u_{\mu} ={~\psi_{\mu} \over \sqrt{-\psi_{\mu} \psi^{\mu}}}\, .
\ee
In the case of interest, namely the isothermal S-brane, $u_{\mu}$ is specified in terms of the thermal potential $\psi$ used to define the temperature:
$$
T=\sqrt {-\psi_{\mu}\psi^{\mu}}~,
$$
where $\psi_{\mu}$ is given by
\be
\psi_{\mu}= \partial_{\mu}\psi \,  ~\Longrightarrow ~~~\partial_{\mu }\psi ~ dx^{\mu}= d\psi\, .
\ee
Hence we can write the metric as follows
\be
ds^2=-N^2 d {\psi}^{2}  +g_{ij}dx^{i}dx^{j} \, ~~\Longrightarrow~~X^0=\psi \, ,
\ee
showing that $X^0$ (which defines the embedding of the S-brane) can be identified with the thermal potential $\psi$. Utilizing this information, the S-brane determinant takes the following form:
\be
{\rm det}(\gamma_{ij})=-{\rm det}(g_{\mu\nu})\left(-g^{\mu \nu}\partial_{\mu}\psi \partial _{\nu} \psi \right )=
-{\rm det}(g_{\mu\nu})~T^2\, .
\ee

We are now in a position to express the action of the isothermal S-brane in very elegant and suggestive form:
\begin{eqnarray}
S_B&=&-\int d\tau d^3x \, \kappa \, e^{-2\phi} \sqrt{-{\rm det}(\tilde g_{\mu\nu})}~\left(\sqrt{-\partial_{\mu} \psi\partial^{\mu} \psi}\right) \delta(\psi-\psi_c)\nonumber \\
\label{SBpsi} &=&-\int d\tau d^3x \, \kappa \, e^{-2\phi} \sqrt{-\tilde g}~\tilde T \delta(\psi-\psi_c)\, , 
\end{eqnarray}
where $\psi_c\equiv \psi(\pi(\vec x),\vec x)$ involves the time  location $\pi(\vec x)$ of the brane. Written in the Einstein frame,
\be
\label{SBpsiE}
S_B=- \int d\tau d^3x  \, \kappa \, e^{\phi} \sqrt{-g}~ T \delta(\psi-\psi_c) .
\ee
The brane action is manifestly re-parametrization invariant in the {\it four dimensional sense}.

\subsection{The S-brane fluctuations}

Although the analysis that follows can be done in full generality, we restrict ourselves to the case of the long wavelength approximation, where we neglect spatial gradients. In this approximation, the relevant fluctuations are given in terms of the  temporal displacement $\pi$ of the brane, as well as the dilaton and $\kappa$ fluctuations. 
It is important to notice that there is no contribution from the S-brane to the energy density even in the presence of fluctuations:
$\delta \rho_B=0$ . This follows from the fact that for the S-brane system $\gamma_B=0$ so that 
\be
 \tilde \rho_B=\gamma_B\, \tilde p_B=0 ~~~~~{\rm and}~~~~~ \rho_B=\gamma_B\,p_B=0 .
\ee
Furthermore, assuming a (block-)diagonal metric and in the long-wavelength approximation the off-diagonal terms of the stress tensor can be set to zero. 
The remaining equations of motion are the one for dilaton $\phi$, 
the transverse part of the metric (parametrized in the string frame by the scale factor $\tilde A$  and in the Einstein frame by $A=\tilde A e^{-\phi}$) 
and that of $\psi$. Concerning the $\phi$ and $\tilde A$ equations, the pre-factor $\tilde A^3 e^{-2\phi}$ is divided out; 
e.g. it has been considered just before and after the locus of the brane. 
Thanks to the continuity conditions for $\tilde A$ and $\phi$, the impact of their fluctuations is already considered 
in the bulk just before and after the S brane. 

The remaining terms needed to be specified are the ratios $\tilde T(\pi)/  \tilde T_c$ and $\kappa/ \kappa(0)$. 
The first ratio is fixed to be unity by the definition of the isothermal S-brane: $\tilde T(\pi)/  \tilde T_c\equiv 1$, 
and determines the locus of the S-brane $\pi$. 
As shown in \cite{KPT}, the brane tension is determined in terms of the extra stringy degrees freedom appearing 
when the temperature reaches its maximal, critical value $T_c$.  At this point $\kappa$ is given by:
\be
\kappa(\pi)= 2\sqrt{6\Lambda}~\tilde T^{2}~e^{\phi(\pi)}\, =\kappa(0)\,e^{\phi(\pi)-\phi(0)}\, ,
\ee
where $\kappa(0)=\kappa_c$ is the tension associated with the background S-brane. 
Expanding around $\pi$, and using Eq. (\ref{Ca}), 
we obtain the shift of $\kappa$ in terms of the metric and dilaton fluctuations: 
\be
\kappa(\pi)=\kappa_c\left(1+\p- C_a \right) \Big \abs_{\pi}~=\kappa_c\left(1+C_3 \right).
\ee

\subsection{Entropy conservation in the combined system $S_T+S_B$}
\label{A4}

The remaining constraint concerning the S-brane comes from the equation of motion of the temperature potential 
$\psi$. Away from the S-brane the thermal entropy is conserved. Also the temperature fluctuations obey the state equation of the background 
\be
\tilde \rho_T+\delta \tilde \rho_T-3(\tilde p_T+\delta \tilde p_T ) \Big \abs_ {\mp} =0,    
\ee
before and after the S-brane. 
We would like to investigate the possibility of entropy production after crossing the S-brane. 
This in principle can happen since the state equation of the brane is different: $\rho_B=0$. 
To answer this question, we consider the equation of motion of $\psi$:  
\be
{ \delta (S_T +S_B)\over\delta \psi}=- 4\Lambda \partial_{\mu}  (\tilde A^3 \tilde T^3 u^{\mu}) + \kappa \Big (\partial_{\mu}( \tilde A^3 e^{-2\phi} \Delta\, u^{\mu} )
-u^{\mu} \tilde A^3 e^{-2\phi}\, \partial_{\mu}\Delta \Big )=0, 
\ee
where we replace the $\delta$-function by one of its smooth representations, $\Delta({\psi})$, in order 
to treat possible ambiguities that may arise from the variation of the $\delta$-function distribution. 
We have also used 
$$
\partial_{\mu}\psi {d \Delta \over d \psi}=\partial_{\mu} \Delta \, ~~~~~{\rm and} ~~~~u^{\mu}={\partial_{\mu} \psi \over \sqrt{-\partial_{\nu}\psi
 \partial^{\nu} \psi}}.
$$
Therefore, the equation of motion of $\psi$ gives rise to a global entropy conservation law, concerning the combined system:
\be
{ \delta (S_T +S_B)\over\delta \psi}=- 4\Lambda \partial_{\mu}  (\tilde A^3 \tilde T^3 u^{\mu}) +
 \kappa \Delta \partial_{\mu}( \tilde A^3 e^{-2\phi}  u^{\mu} )=0.
\ee

The equation above can be integrated in the interval $(\pi_-=\pi-\epsilon ,~\pi_+=\pi+\epsilon)$, where $\epsilon$ is the width  of $\Delta$. 
In the long wavelength approximation where the velocity vector becomes $u^{\mu}=(1,0,0,0)$, we obtain:
\be
\label{Gentropy}
\left(4\Lambda \tilde A^3 \tilde T^3 \right)_{\pi_-} -\left(4\Lambda \tilde A^3 \tilde T^3 \right)_{\pi_+}=-{\kappa\over 2\,\tilde T_c}{ \left[  \left(\tilde A^3 e^{-2\phi}\right)_{\pi_-}^{\prime}+\left(\tilde A^3 e^{-2\phi}\right)_{\pi_+}^{\prime} \right] }\, .
\ee
The LHS gives the difference of the entropy in the contracting phase (before the S-brane) and the entropy in the expanding phase (after the brane). The RHS 
is associated with the entropy change produced by the S-brane. Utilizing the relation of the entropy with the background entropy we have:
\be
\left(4\Lambda \tilde A^3 \tilde T^3 \right)_{\pi_-} -\left(4\Lambda \tilde A^3 \tilde T^3 \right)_{\pi_+}=4\Lambda \tilde a_c^3 \tilde T_c^3 \left(\, 3c_{\sigma}^--3c_{\sigma}^+\right)\, .
\ee
 Therefore the LHS is proportional to the difference between the coefficients $c_{\sigma}^--c_{\sigma}^+$. 
However, the existence of a unique maximal temperature $\tilde T_c$ together with the continuity condition $\tilde A(\pi_-)=\tilde A(\pi_+)$ imply that 
 $c_{\sigma}^-=c_{\sigma}^+$. In other words, there is no entropy production by the S-brane. 
Taking into account that $\tilde A^{\prime}(\pi_-)=\tilde A^{\prime}(\pi_+)=0$, the  brane contribution to the entropy becomes
 \be
 -{\kappa\over 2\,\tilde T_c}{ \left[  \left(\tilde A^3 e^{-2\phi}\right)_{\pi_-}^{\prime}+\left(\tilde A^3 e^{-2\phi}\right)_{\pi_+}^{\prime} \right] }= {\kappa \over \,\tilde T_c}{ \tilde A^3 \big[\phi^{\prime}(\pi_-)+ \phi^{\prime}(\pi_+) \big] }\,.
 \ee  
 Therefore, the vanishing of the contribution of the S-brane to the entropy implies reflecting conditions for $\phi^{\prime}$, and the equality of $c_{\p}^\pm$
 and $C_{3 }^\pm$ -- the latter follows from the continuity of $\phi$ across the brane:
 \be
 \label{reflect}
 {\kappa \over \,\tilde T_c}{ \tilde A^3 \big[\phi^{\prime}(\pi_-)+ \phi^{\prime}(\pi_+) \big] }= {\kappa \over \,\tilde T_c}{ \tilde A^3 {\sqrt{3}\over2\abs \xi\abs}}(c_{\p}^--c_{\p}^+)=0\, .
 \ee

\section{Derivation of $\Phi$ and $\p$}
\label{B}

The goal of this appendix is to derive the analytic solutions for the metric and dilaton fluctuations $\Phi$ and $\p$, Eqs (\ref{solphi}) and (\ref{solp}) in the long wavelength approximation. The starting point is the second order differential equation (\ref{peom}) for $\Phi$, where $\epsilon= {\rm sign}\, \tau$ and $y$ is defined in Eq. (\ref{y}).
Expressing $\Phi$ in terms of the function $F$,
\eq{defF}{\Phi \equiv \frac{F}{y^2} - \frac{c_\p}{4}, }
and utilizing the explicit expressions (\ref{hsol}) and (\ref{hpsol}) for $\mathcal H$ and $\mathcal H^{\prime}$, we obtain the following equation for $F$:
\eq{EqF}{ F^{\prime\prime}-2\left(\frac{1}{(\tau+\epsilon \xi_+)^2}+\frac{1}{(\tau+\epsilon \xi_-)^2} -\frac{1}{(\tau+\epsilon \xi_+)(\tau+\epsilon \xi_-)}\right)F=0.}
Notice that the expression multiplying $F$ is nothing but the ratio $\mathcal H^{\prime \prime}/ \mathcal H$, and so,
\eq{EqF2}{ F^{\prime\prime}-\frac{\mathcal H^{\prime \prime}}{ \mathcal H}F=0,}
implying the particular solution $F=\hat C_1 \mathcal H$, where  $\hat C_1$ is an integration constant. To obtain the full solution, we set $F\equiv \mathcal H f$ 
giving us the following simple equation for $f$:
\eq{Eqf}{ f^{\prime \prime} \mathcal H+2f^{\prime} \mathcal H^{\prime}=0\quad \Longrightarrow\quad  f^{\prime}= \frac{3C_2}{{\mathcal H^{2}}} \quad \Longrightarrow\quad f=3C_2\int {d\tau\over \mathcal H^2}\, .}
The last integral  can be evaluated analytically. To this extend, it is convenient to convert it into  a $y$-integral,
\be
\label{int}
\int d\tau \, (\cdots )=\int {dy\over y\mathcal H}\, (\cdots )=\epsilon \int dy \, {y \over \sqrt{y^2 +\xi^2}}\, (\cdots) \, ,
\ee
and use the expression of $\mathcal H$ in terms of $y$. The integration is straightforward and leads 
\be
f=3C_2\epsilon \left( \frac{1}{3}(y^2-2\xi^2)\sqrt{y^2+\xi^2}  - \xi^2 {y^2+2\xi^2 \over \sqrt{y^2+\xi^2} } \right)+\hat C_1={ C_2 y^2 \over \mathcal H} \left(1+4\xi^2\mathcal H^{\prime} \right)+\hat C_1\, .
\ee
Finally the full expression for $\Phi$ reads:
\eq{2sol}{\Phi = \hat C_1 \frac{\mathcal H}{y^2} + C_2  (1+ 4\xi^2 \mathcal H^{\prime}\,) -\frac{c_\p}{4}~. }
Furthermore, utilizing the relation between the dilaton fluctuation and $\Phi$, Eq. (\ref{solpp}),  we obtain:
\eq{dilatonfluctuation1}{\p^{\prime} = \phi_0^{\prime}\, 4\left( \hat C_1 \frac{\mathcal H}{y^2} + C_2  (1+ 4\xi^2 \mathcal H^{\prime}\,)  \right)\,.}
To find $\varphi$ analytically, it is convenient to use the expressions (\ref{psol})--(\ref{hpsol}) for 
$\phi_0^{\prime}$, $\mathcal H$ and $\mathcal H'$ in terms of $y$, and convert the integral as in Eq. (\ref{int}) to get
\begin{eqnarray}
\p \!\!&=&\!\! -\epsilon 4\sqrt{3}\xi {\hat C_1} \int {dy\over y^5} +4C_2 \phi_0 
+16C_2\sqrt{3} \xi^3 \int {dy\over y^5} {y^2+2\xi^2 \over \sqrt{y^2+\xi^2}} \nonumber\\
\!\!&=&\!\!-\hat C_1 {\phi_0^{\prime}\over y^2} +4C_2 \phi_0  
+16C_2\sqrt{3}   \left(   {\xi  \sqrt{y^2+\xi^2} \over 4 y^2}  
- {\xi^3 \sqrt{y^2+\xi^2}  \over 2y^4 }+{1\over 8 }L(y)   \right)+{\rm cst}
\end{eqnarray}
where
$$
L(y)= \ln{\sqrt{y^2+\xi^2}-\xi \over \sqrt{y^2+\xi^2}+\xi}= \ln{ \abs\tau\abs+ \xi_- \over  \abs\tau\abs+\xi_+}=-{2\over \sqrt{3}}\,  \phi_0+ {\rm cst}\,,
$$
as follows from the background expression (\ref{background}) for $\phi_0$. Therefore, the term proportional to $\phi_0$ cancels and we  obtain
\be
\label{sol3}
\p =- \hat C_1 {\phi_0^{\prime}\over y^2} 
-4C_2\phi_0 ^{\prime} \mathcal H (y^2-2\xi^2)  +C_{3}\, .
\ee
In the expressions (\ref{2sol}) and (\ref{sol3}) for $\Phi$ and $\varphi$, it is more appropriate to replace the terms proportional to $\hat C_1/ y^2$ 
with terms proportional to $\phi_0^{\prime}$. It is also convenient to introduce the dimensionless integration constant $C_1$ by a suitable rescaling 
of $\hat C_1$ by $\xi$. Doing so we obtain the final expressions (\ref{solphi}) and (\ref{solp}) for $\Phi$ and $\p$ in terms of the two fundamental functions characterizing the background, 
$\mathcal H$ and $\phi^{\prime}_0$.

\section{Equidilaton and equipotential gauges}
\label{adm4}

For conceptual clarity (and with an eye on follow up investigations) one can consider deriving our results in two other gauges that are naturally suggested by the relevant degrees of freedom of the system. The first gauge in question gauges away the dilaton perturbation (``equidilaton gauge") and the other gauges away the thermal potential (``equipotential gauge"). Although similar in spirit, \textit{neither gauge is equivalent to the more familiar comoving gauge of single field scalar cosmology}, as the two fluid nature of the system will always ensure a non-vanishing momentum flux for a comoving observer.

We begin by making use of the ADM decomposition
\eq{adm}{ds^2 = -N^2dt^2 + h_{ij}(dx^i + N^idt)(dx^j + N^jdt).}
Away from the S-brane, we can express the action of the string gas/dilaton fluid as the sum of:
\eq{g}{S_G = \frac{1}{2}\int d^4x \sqrt{h}\left[N R^{(3)} + \frac{1}{N}(E^{ij}E_{ij} - E^2)\right]}
\eq{d}{S_\phi = \int d^4x \sqrt{h}\left[\frac{1}{N}(\dot\phi - N^i\partial_i\phi)^2 - Nh^{ij}\partial_i\phi\partial_j\phi\right]}
\eq{sg}{S_\psi = \Lambda\int d^4x \sqrt{h}N\left[\frac{1}{N^2}(\dot\psi - N^i\partial_i\psi)^2 - h^{ij}\partial_i\psi\partial_j\psi\right]^2}
where
\eq{edef}{E_{ij} = N K_{ij} := \frac{1}{2}\left[\dot h_{ij} - \nabla_iN_j - \nabla_jN_i\right],}
$K_{ij}$ is the extrinsic curvature of the foliation, and where indices are raised/lowered and covariant derivatives defined with respect to the 3-metric $h_{ij}$.

Once we have fixed a gauge, we can solve for the shift function and the lapse vector and substitute back into the action to obtain the action for the perturbations to a given order. One only needs to solve for the constraints up to linear order in the perturbations to obtain the action up to cubic order. This is because terms in the action that come from solving the constraints to cubic order multiply the zeroth order constraint equations (which vanish by the equations of motion), and those that come from solving the constraints to quadratic order multiply the first order solutions to the constraint equations, and so also vanish \cite{jm}.


\subsection{Equidilaton gauge}

Equidilaton gauge is defined as having foliated spacetime such that the dilaton fluctuations have completely been gauged away. That is\footnote{Note that there is no anisotropic stress in our system to linear order in perturbation theory.}:
\bea
\phi(t,x) &\!\!=\!\!& \phi_0(t)\nn\\ \nn
\psi(t,x) &\!\!=\!\!& \psi_0(t) + \sigma(t,x)\\ 
\label{cgf}h_{ij}(t,x) &\!\!=\!\!& a^2(t)e^{2\calR(t,x)}\delta_{ij}\, .
\eea
In this gauge, the momentum and Hamiltonian constraint equations (to first order) become
\eq{mcc}{\nabla_i\left[N^{-1}(E^i_k - E\delta^i_k)\right] = 4\Lambda\frac{\dot\psi_0^3}{N^3}\partial_k\sigma}
\eq{hcc}{\frac{R^{(3)}}{2} - \frac{1}{2N^2}(E^{ij}E_{ij} - E^2) - \frac{\dot\phi_0^2}{N^2} - 3\frac{\Lambda}{N^4}\dot\psi^4= 0.}
Writing
\bea
N &\!\!=\!\!& 1 + \alpha_1 \nn\\ 
\label{exp} N^i &\!\!=\!\!& \partial_i\theta + N^i_T~,~w/ \partial_i N^i_T\equiv 0
\eea
where $\alpha_1, \theta$ and $N^i_T$ are all first order quantities, we find the solutions
\eq{asolc}{\alpha_1 = \frac{\dot \calR}{H} + \frac{2\Lambda}{H}\dot\psi_0^3\sigma}
\eq{tsolc}{\partial^2\theta = -\frac{\partial^2\calR}{a^2 H} + \left\{3\dot\calR\left(1 + \frac{\Lambda\dot\psi_0^4}{H^2}\right) - \frac{6\Lambda\dot\psi_0^4}{H}\left(\frac{\dot\sigma}{\dot\psi_0} - \frac{\Lambda\dot\psi_0^4}{H}\frac{\sigma}{\dot\psi_0}\right)\right\}, }
where $\partial^2 = \partial_i\partial_i$ contains no factors of the scale factor. 
We substitute the above back into the action, liberally integrate by parts and make use of the background equations of motion
\bea
3H^2 = \dot\phi_0^2 + 3\Lambda\dot\psi_0^4~,~\ddot\phi_0 = -3H\dot\phi_0~,~ \ddot\psi_0 = -H\dot\psi_0,
\eea
which results in (after transforming to conformal time)
\bea
S &\!\!=\!\!& \int d^4x\, \left[3\left(a^2 + \frac{\Lambda\psi_0'^4}{\calH^2}\right)\calR'^2  - 3\left(a^2 - \frac{\Lambda\psi_0'^4}{3\calH^2}\right)(\partial \calR)^2\right] \nn\\ \nn &\!\!+\!\!& 6\Lambda\psi_0'^2 \int d^4x~\left[\sigma'^2 - \frac{1}{3}(\partial\sigma)^2\right]\\\label{cact}
&\!\!+\!\!& 2\Lambda\psi_0'^3\int d^4x\left[-6 \frac{\sigma'\calR'}{\calH} + 6\calR'\sigma\frac{\Lambda\psi_0'^4}{a^2\calH^2} + \frac{2}{\calH}\partial\calR\cdot\partial\sigma\right],
\eea
where we extract $\psi'_0$ from the integrand as it is independent of time. The equations of motion that result from the above are:
\eq{req}{\calR'' + 2\calH\calR'\left(1 + \frac{\Lambda\psi_0'^4}{a^2\calH^2}\right) - \partial^2\calR = \frac{4\Lambda\psi_0'^4}{a^2}\left[\frac{\sigma'}{\psi_0'} - \frac{\Lambda\psi_0'^4}{\calH a^2}\frac{\sigma}{\psi_0'}\right]}
\eq{seq}{\sigma'' -\frac{1}{3}\partial^2\sigma = \frac{\psi_0'}{\calH}\left[\calR'' + 2\calH\calR' - \frac{1}{3}\partial^2\calR \right].} 
In the long wavelength limit (\ref{seq}) integrates to
\eq{seqint}{\frac{\calR'}{\calH} = \left[\frac{\sigma'}{\psi_0'} -\frac{\Lambda\psi_0'^4}{\calH a^2}\frac{\sigma}{\psi'_0} \right] + \frac{K_{ed}}{\calH a^2}}
where $K_{ed}$ is an integration constant (the subscript denotes the gauge in which we have defined it) and where we have used the fact that
\eq{a2e}{a^2\calH = \Lambda\psi_0'^4\tau + \calH_c a_c^2\, ,}
which follows from $(\calH a^2)' = \Lambda\psi_0'^4$ and determining the integration constant through background quantities at the critical temperature. In order to see the existence of two independent solutions for $\calR$, we substitute the above into (\ref{req}) to result in
\eq{reqi}{\calR'' + 2\calH\calR'\left(1 - \frac{\Lambda\psi_0'^4}{a^2\calH^2}\right) + \frac{4\Lambda\psi_0'^4}{\calH a^4}K_{ed} = 0.}

\subsection{Equipotential gauge}

Equipotential gauge is defined by the foliation that gauges away all thermal potential fluctuations:
\bea
\phi(t,x) &\!\!=\!\!& \phi_0(t) + \p(t,x)\nn\\ \nn
\psi(t,x) &\!\!=\!\!& \psi_0(t)\\ 
\label{eqgf} h_{ij}(t,x) &\!\!=\!\!& a^2(t)e^{2\tilde\calR(t,x)}\delta_{ij},
\eea
where we distinguish the variable $\tilde\calR$ from the variable $\calR$ in equidilaton gauge, as they are defined on different hypersurfaces. In this gauge, the momentum and Hamiltonian constraint equations become
\eq{mceq}{\nabla_i\left[N^{-1}(E^i_k - E\delta^i_k)\right] = 2\frac{\dot\phi_0}{N}\partial_k\p}
\eq{hceq}{\frac{R^{(3)}}{2} - \frac{1}{2N^2}(E^{ij}E_{ij} - E^2) - \frac{\dot\phi^2}{N^2} - 3\frac{\Lambda}{N^4}\dot\psi_0^4= 0}

\eq{asoleq}{\alpha_1 = \frac{\dot{\tilde\calR}}{H} + \frac{\dot\phi_0}{H}\p}
\eq{tsoleq}{\partial^2\theta = -\frac{\partial^2\tilde\calR}{a^2 H} + \left\{3\dot{\tilde\calR}\left(1 + \frac{\Lambda\dot\psi_0^4}{H^2}\right) - \frac{\dot\phi_0}{H}\dot\p + 3\Lambda\dot\phi_0\frac{\dot\psi_0^4}{H^2}\p \right\} .}
The quadratic action that results upon solving for the constraints is now 
\bea
S &\!\!=\!\!& \int d^4x\, \left[3\left(a^2 + \frac{\Lambda\psi_0'^4}{\calH^2}\right)\tilde \calR'^2  - 3\left(a^2 - \frac{\Lambda\psi_0'^4}{3\calH^2}\right)(\partial \tilde \calR)^2\right]\nn\\ \nn &\!\!+\!\!& \int d^4x~a^2\left[\p'^2 - (\partial\p)^2 + \frac{2\Lambda\psi_0'^4}{a^2\calH^2}\phi_0'^2\p^2\right]\\ 
\label{eqact} &\!\!+\!\!& \int d^4x~\phi_0'a^2\left[-2 \frac{\p'\tilde\calR'}{\calH} + 6\tilde\calR'\p\frac{\Lambda\psi_0'^4}{a^2\calH^2} + \frac{2}{\calH}\partial\tilde\calR\cdot\partial\p\right].
\eea
The equations of motion that result are
\eq{reqeq}{\tilde\calR'' + 2\calH\tilde\calR' - \frac{1}{3}\partial^2\tilde\calR = -\frac{2}{3}\phi_0'\p' }
\eq{peqeq}{\p'' +2\calH\p'-\partial^2\p - 2\Lambda\psi_0'^4\frac{\phi_0'^2}{a^2\calH^2}\p = \frac{\p_0'}{\calH}\left[\tilde\calR'' + 2\calH\tilde\calR'\left(1 + \frac{\Lambda\psi_0'^4}{a^2\calH^2}\right) - \partial^2\tilde\calR \right].} 
In the long wavelength limit, we can immediately integrate (\ref{reqeq}):
\eq{rinteq}{\tilde\calR' = -\frac{2}{3}\phi_0'(\p + K_{ep})\, , }
which implies that (\ref{peqeq}) becomes
\eq{peqeq2}{\p'' + 2\calH\p'\left(1+ \frac{\phi_0'^2}{3\calH^2}\right)=\frac{2}{3}
\frac{\phi_0'^2\Lambda\psi_0'^4}{a^2\calH^2}(\p-2K_{ep})\, , }
which can be written as
\eq{peqeq3}{\p'' -2\frac{\calH'}{\calH}\p' = \frac{2}{3}
\frac{\phi_0'^2\Lambda\psi_0'^4}{a^2\calH^2}(\p-2K_{ep})\, .}

\subsection{Relating the gauges and the meaning of isocurvature perturbations}

One might ask, how these two gauges relate? In equidilaton gauge, we work in the slicing where all dilaton fluctuations have been gauged away:
\eq{cmgdf}{\phi(t,x) = \phi_0(t)~,~\psi(t,x) = \psi_0 + \sigma(t,x)}
whereas in equipotential gauge we work in a slicing where all thermal potential fluctuations have been gauged away:
\eq{eqgdf}{\phi(t,x) = \phi_0(t) + \p(t,x)~,~\psi(t,x) = \psi_0\, .}
Clearly the two slicings can be related by an infinitessimal time reparametrization
\eq{trp}{\tilde t = t + \Pi}
where the tilde'd time coordinate is that of equipotential gauge. From the transformation of the metric under such a reparametrization, it is straightforward to show that starting in comoving gauge with perturbations $\cal R$ and $\sigma$ that under (\ref{trp}):
\bea
\label{rtfl}
\calR&\to& \calR -\calH\Pi\\
\sigma&\to&\sigma - \psi_0'\Pi\\
\label{df}
\p&\to& -\phi_0'\Pi\, , 
\eea
where the latter follows from the fact that we have now moved out of the slicing where the fluctuations of $\phi$ vanish (in equidilaton gauge $\p \equiv 0$). If we identify this new slicing with the equipotential slicing -- where $\sigma$ is gauged away to vanish -- then this defines $\Pi$ as
\eq{pidef}{\Pi = \frac{\sigma}{\psi_0'}.}
Therefore the dilaton fluctuation in this gauge is given by (\ref{df}):
\eq{isom}{\p = -\frac{\phi_0'}{\psi_0'}\sigma}
and hence
\eq{isocd}{\p =  -\frac{d\phi_0}{d\psi_0}\sigma\, .}
This expression highlights the physical meaning of an isocurvature perturbation -- when a fluid has two (or more) fluctuating components, if their background solutions are both monotonic (i.e. they define equally good clocks), and if their parametric dependence on time is not proportional to a constant, then the two describe independent fluctuations and are thus non-adiabatic.

In addition, we can also perform an important consistency check -- the action (\ref{cact}) must therefore relate to the action (\ref{eqact}) via the field redefinition
\bea
\calR &\!\!=\!\!& \tilde\calR - \frac{\calH}{\phi_0'}\p\\
\sigma &\!\!=\!\!& -\frac{\psi_0'}{\phi_0'}\p\, , 
\eea
which follows from Eqs (\ref{rtfl}) to (\ref{isom}). It is straightforward though tedious to check that this is indeed true.

\subsection{Relation to the Mukhanov-Sasaki variable}

In both gauges, we note that the quadratic actions for $\calR$ and $\tilde\calR$ are identical, with isomorphic interaction terms (the last lines of (\ref{cact}) and (\ref{eqact})). Since in both cases, the corresponding variable $\calR$ or $\tilde\calR$ is conserved in the absence of an isocurvature component, we see that they both relate to the adiabatic mode. From this, we can easily construct the canonically normalized action for the Mukhanov-Sasaki variable \cite{Sasaki, Mukh}:
\eq{qactg}{S^{(2)}_G = \frac{1}{2}\int~d^4x~ \left[v'^2 - c_s^2(\partial v)^2 + \frac{z''}{z}v^2\right]}
with
\eq{vdef}{v = z\calR,~\text{or}~v = z\tilde\calR} 
depending on the slicing, and where
\eq{zdef}{z = a\sqrt{6}\left(1 + \frac{\Lambda\psi_0'^4}{a^2\calH^2}\right)^{1/2} = \frac{a\phi_0'}{\calH}\left(2 + 6\frac{\Lambda\psi_0'^4}{\phi_0'^2a^2}\right)^{1/2} }
on either foliation. The latter relation compares familiarly with the analogous expression for a single minimally coupled scalar field ($z = a\phi_0'/\calH$).  The speed of sound is given by
\eq{csdef}{c_s^2 = \frac{1 - \frac{\Lambda\psi_0'^4}{3a^2\calH^2}}{1 + \frac{\Lambda\psi_0'^4}{a^2\calH^2}}\, , }
which clearly interpolates between $1/3 \leq c^2_s \leq 1$ as the evolution moves from thermal string gas domination ($\Lambda\psi_0'^4/a^2\calH^2 \to 1$) to dilaton domination $\Lambda\psi_0'^4/a^2\calH^2 \to \Lambda\psi_c'^4/a_c^2\calH_c^2$, where the latter can be made vanishingly small depending on the parameters of the problem (the total entropy of the universe).

We can readily cast the second order action in terms of two canonically normalized variables. From (\ref{cact}), we see that $\sigma$ is only a trivial rescaling away from being canonically normalized. Using (\ref{vdef}) and (\ref{zdef}), and defining
\eq{mudef}{u := \mu \frac{\sigma}{\psi_0'}~;~\mu = \sqrt{12\Lambda}\psi_0'^2\, ,} 
the canonically normalized action in equidilaton gauge is given by
\bea
S &\!\!=\!\!&  \frac{1}{2}\int~d^4x~ \left[v'^2 - c_s^2(\partial v)^2 + \frac{z''}{z}v^2\right]\nn\\ \nn
&\!\!+\!\!& \frac{1}{2}\int~d^4x~ \left[u'^2 - \frac{1}{3}(\partial u)^2\right]\\
\label{can2f}
&\!\!+\!\!& \mu\int~d^4x~ \left[-\frac{u'}{\calH}\left(\frac{v}{z}\right)' + u\frac{\Lambda\psi_0'^4}{a^2\calH^2}\left(\frac{v}{z}\right)' + \frac{1}{3\calH z}\partial u\cdot\partial v\right].
\eea

\subsection{Relation to quantities in longitudinal gauge}

In order to understand the results of the calculations in the main body of the paper performed in longitudinal gauge, it is instructive to explicitly relate quantities in this gauge to either of the gauges considered in this appendix. As exact expressions for the mode functions are easiest to obtain in equidilaton gauge, we flesh out the relation between these two gauges. We assert that the time reparametrization $t \to t + \xi^0$, with
\eq{gtt}{\xi^0 = -a\theta}
where $\theta$ is defined by (\ref{tsolc}), such that
\begin{eqnarray}
a\theta &\!\!=\!\!& -\frac{\calR}{\calH} + \partial^{-2}\left\{3\calR'\left(1 + \frac{\Lambda\psi_0^{'4}}{a^2\calH^2}\right) - \frac{6\Lambda\psi_0^{'4}}{a^2\calH}\left(\frac{\sigma'}{\psi'_0} - \frac{\Lambda\psi_0^{'4}}{a^2\calH}\frac{\sigma}{\psi'_0}\right)\right\}\nonumber \\ &\!\!:=\!\!& -\frac{\calR}{\calH} + \partial^{-2}\chi\, ,\label{tsolcc}
\end{eqnarray}
transforms equidilaton gauge into longitudinal gauge
\eq{}{ds^2 \, = \, -a^2(1+2\Phi)d\tau^2 + a^2(1- 2\Phi)dx^2 \, ,}
where the gravitational potential $\Phi$ is given by
\eq{longdef}{\Phi = -\calH\partial^{-2}\chi\, .}
Moreover, it is informative to see how the various perturbed quantities in equidilaton gauge transformed under this gauge transformation. By (\ref{tsolcc}) and (\ref{longdef}) we see that 
\eq{atans}{a\theta = -\frac{\calR}{\calH} - \frac{\Phi}{\calH}\, ,}
and from (\ref{gtt}), it is clear that moving out of equidilaton gauge, the scalar field now has a fluctuation
\eq{sff}{\p = -\phi_0'\xi^0 = a\theta\phi_0' = -\frac{\calR\phi'_0}{\calH} - \frac{\Phi\phi'_0}{\calH}}
or that the equidilaton curvature perturbation is given by longitudinal gauge quantities as
\eq{rlg}{-\calR = \Phi + \frac{\calH\p}{\phi_0'}\, ,}
where the minus sign in front of $\calR$ is because of the convention defined in (\ref{cgf}). Furthermore, the thermal potential perturbation transforms as
\eq{tpt}{\sigma_l = \sigma_{ed} -\psi_0'\xi^0 = \sigma_{ed} + \psi_0'a\theta\, ,}
and so
\eq{tlg}{\frac{\sigma_{ed}}{\psi_0'} = \frac{\sigma_l}{\psi_0'} - \frac{\p}{\phi_0'}\, ,}
where the subscripts $ed$ and $l$ denote the thermal potential fluctuation in equidilaton and longitudinal gauges respectively. It is straightforward to check that under the identifications (\ref{rlg}) and (\ref{tlg}) that all of the equations of motion for the perturbations $\Phi, \sigma_l$ and $\p$ (\ref{00})--(\ref{vpp2}) follow from (\ref{req}), (\ref{seq}) and the background equations of motion.

\end{document}